\documentclass[runningheads]{llncs}
\usepackage{graphicx}

\usepackage{tikz}
\usepackage{comment}
\usepackage{amsmath,amssymb} 
\usepackage{color}

\newsavebox\CBox
\def\textBF#1{\sbox\CBox{#1}\resizebox{\wd\CBox}{\ht\CBox}{\textbf{#1}}}

\newcommand{\eg}{{\emph{e.g.}}}
\newcommand{\ie}{{\emph{i.e.}}}

\newcommand{\etc}{{\emph{etc}\ }}
\definecolor{grayhighlight}{RGB}{213,229,255}

\begin{document}
\pagestyle{headings}
\mainmatter
\def\ECCVSubNumber{100}  

\title{Realistic Bokeh Effect Rendering on Mobile GPUs, Mobile AI \& AIM 2022 challenge: Report} 

\titlerunning{ECCV-22 submission ID \ECCVSubNumber}
\authorrunning{ECCV-22 submission ID \ECCVSubNumber}
\author{Anonymous ECCV submission}
\institute{Paper ID \ECCVSubNumber}

\author{Andrey Ignatov \and Radu Timofte \and
Jin Zhang \and Feng Zhang \and Gaocheng Yu \and Zhe Ma \and Hongbin Wang \and
Minsu Kwon \and
Haotian Qian \and Wentao Tong \and Pan Mu \and Ziping Wang \and Guangjing Yan \and
Brian Lee \and Lei Fei \and Huaijin Chen \and
Hyebin Cho \and Byeongjun Kwon \and Munchurl Kim \and
Mingyang Qian \and Huixin Ma \and Yanan Li \and Xiaotao Wang \and Lei Lei $^*$
}

\institute{}
\titlerunning{Realistic Bokeh Effect Rendering on Mobile GPUs}
\authorrunning{A. Ignatov, R. Timofte et al.}
\maketitle

\begin{abstract}

As mobile cameras with compact optics are unable to produce a strong bokeh effect, lots of interest is now devoted to deep learning-based solutions for this task. In this Mobile AI challenge, the target was to develop an efficient end-to-end AI-based bokeh effect rendering approach that can run on modern smartphone GPUs using TensorFlow Lite. The participants were provided with a large-scale EBB! bokeh dataset consisting of 5K shallow / wide depth-of-field image pairs captured using the Canon 7D DSLR camera. The runtime of the resulting models was evaluated on the Kirin 9000's Mali GPU that provides excellent acceleration results for the majority of common deep learning ops. A detailed description of all models developed in this challenge is provided in this paper.

\keywords{Mobile AI Challenge, Bokeh, Portrait Photos, Mobile Cameras, Shallow Depth-of-Field, Mobile AI, Deep Learning, AI Benchmark}
\end{abstract}

{\let\thefootnote\relax\footnotetext{%
$^*$Andrey Ignatov \textit{(andrey@vision.ee.ethz.ch)} and Radu Timofte \textit{(radu.timofte@uni-wuerzburg.de)} are the main Mobile AI \& AIM 2022 challenge organizers . The other authors participated in the challenge. \vspace{2mm} \\ Appendix \ref{sec:apd:team} contains the authors' team names and affiliations. \vspace{2mm} \\ Mobile AI 2022 Workshop website: \\ \url{https://ai-benchmark.com/workshops/mai/2022/}
}}

\section{Introduction}

\begin{figure*}[t!]
\centering
\setlength{\tabcolsep}{1pt}
\resizebox{\linewidth}{!}
{
\begin{tabular}{cccccc}
Wide Depth-of-field & Shallow Depth-of-field & Wide Depth-of-field & Shallow Depth-of-field & Wide Depth-of-field & Shallow Depth-of-field\\
    \includegraphics[width=0.34\linewidth]{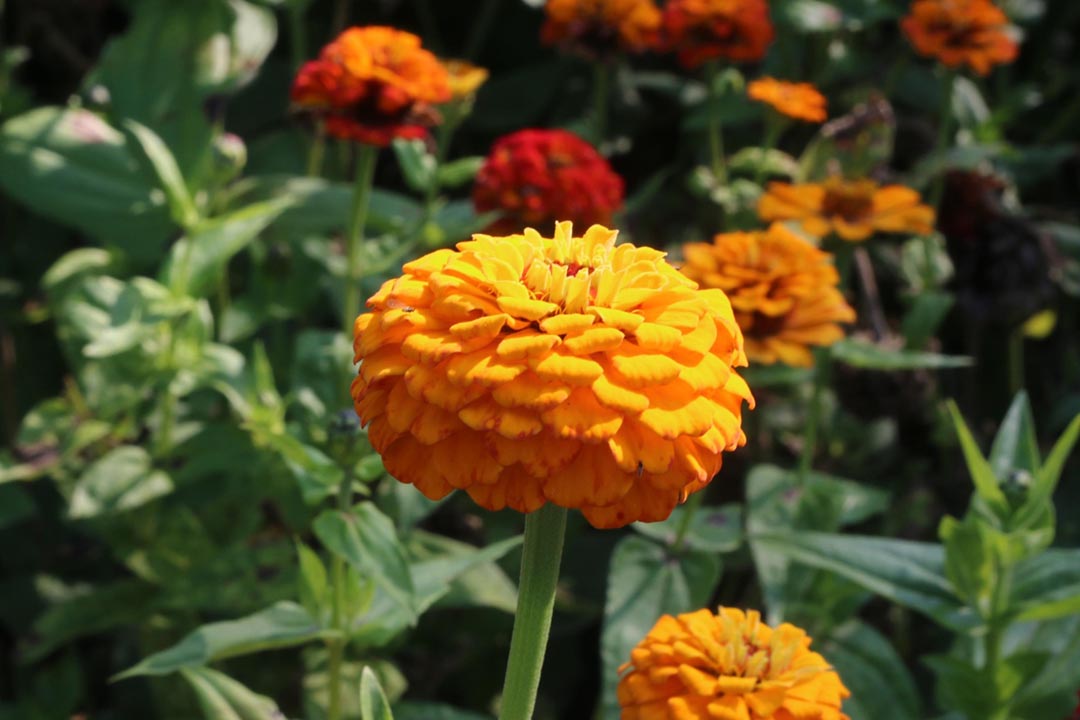}&
    \includegraphics[width=0.34\linewidth]{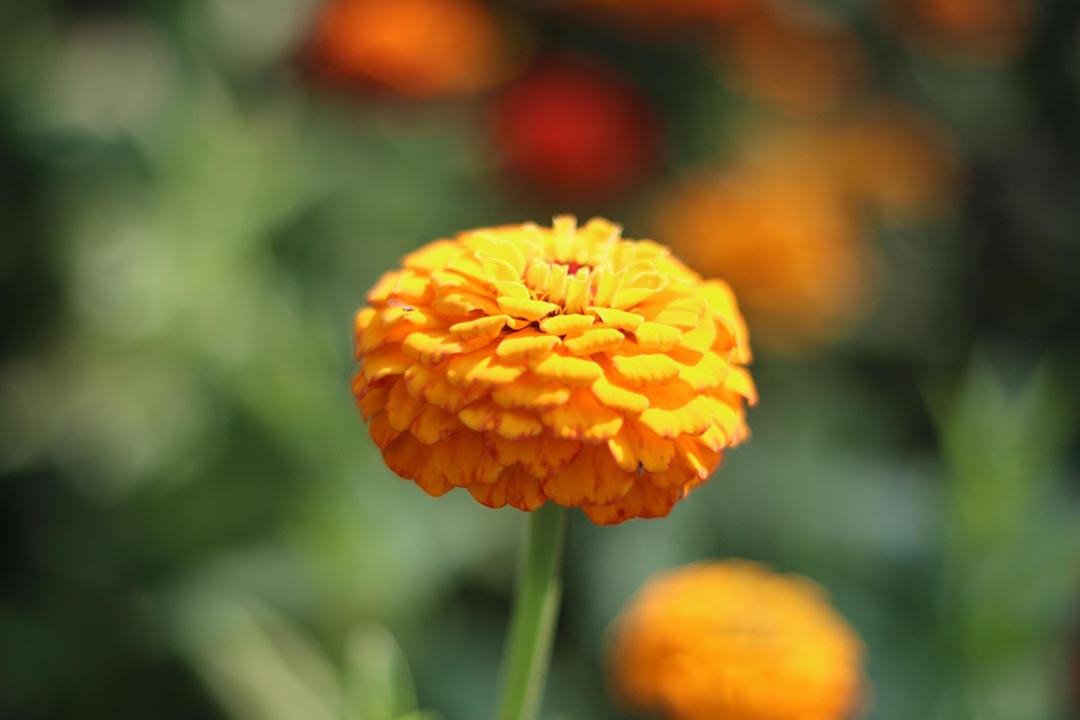}& \hspace{0.8mm}
    \includegraphics[width=0.34\linewidth]{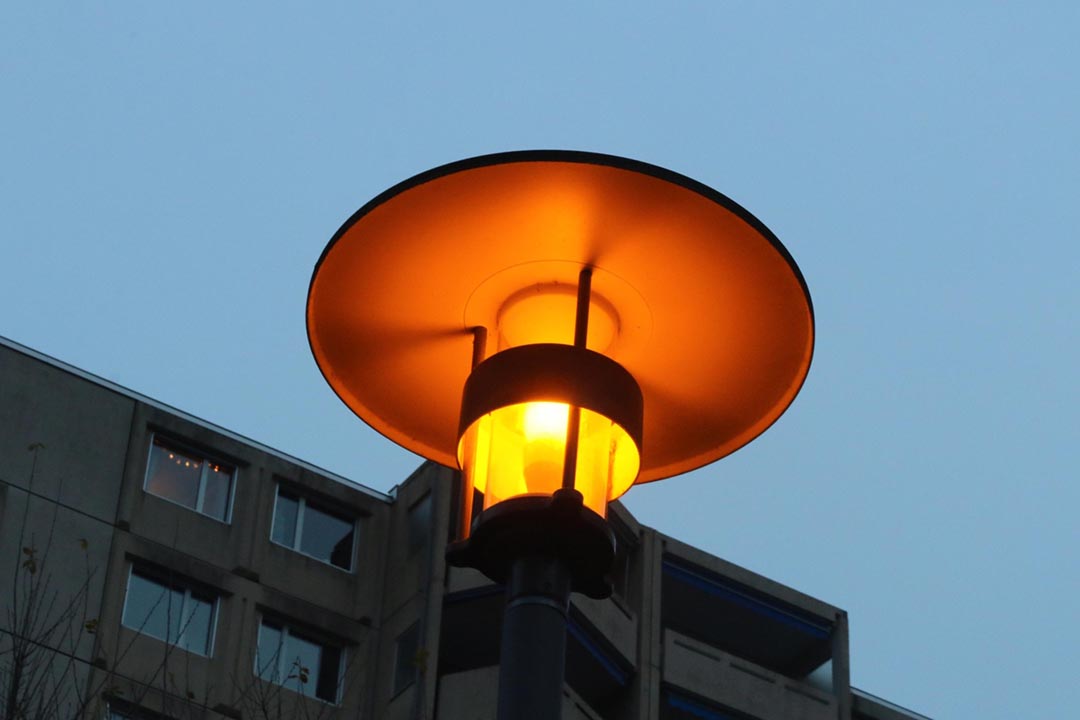}&
    \includegraphics[width=0.34\linewidth]{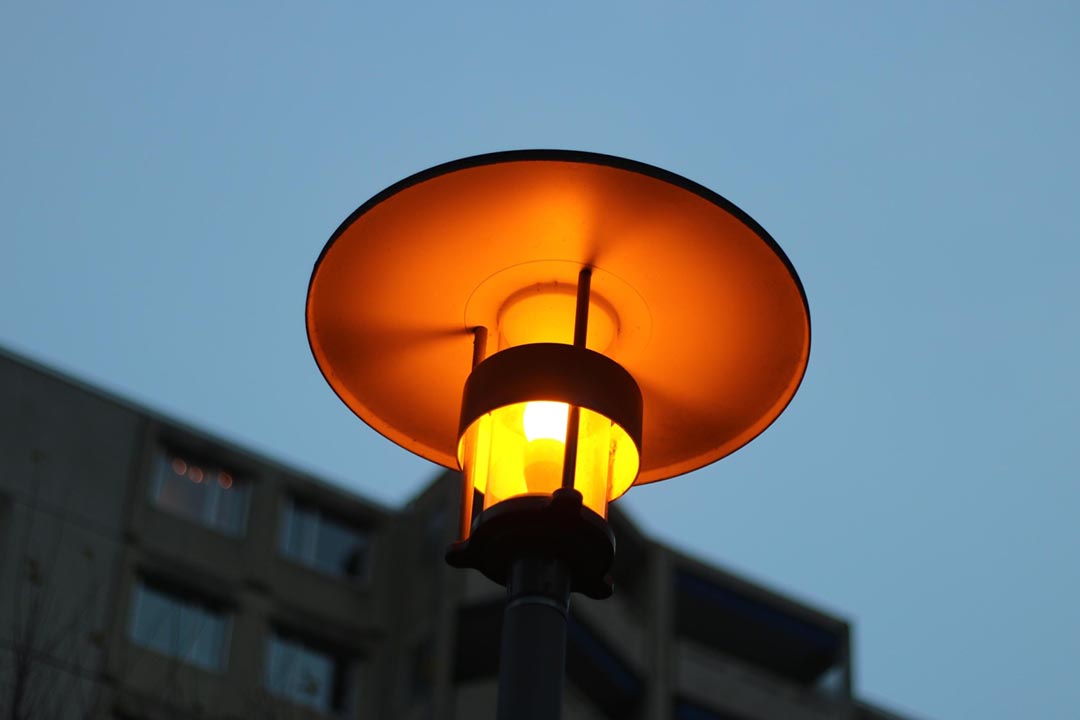}& \hspace{0.8mm}
    \includegraphics[width=0.34\linewidth]{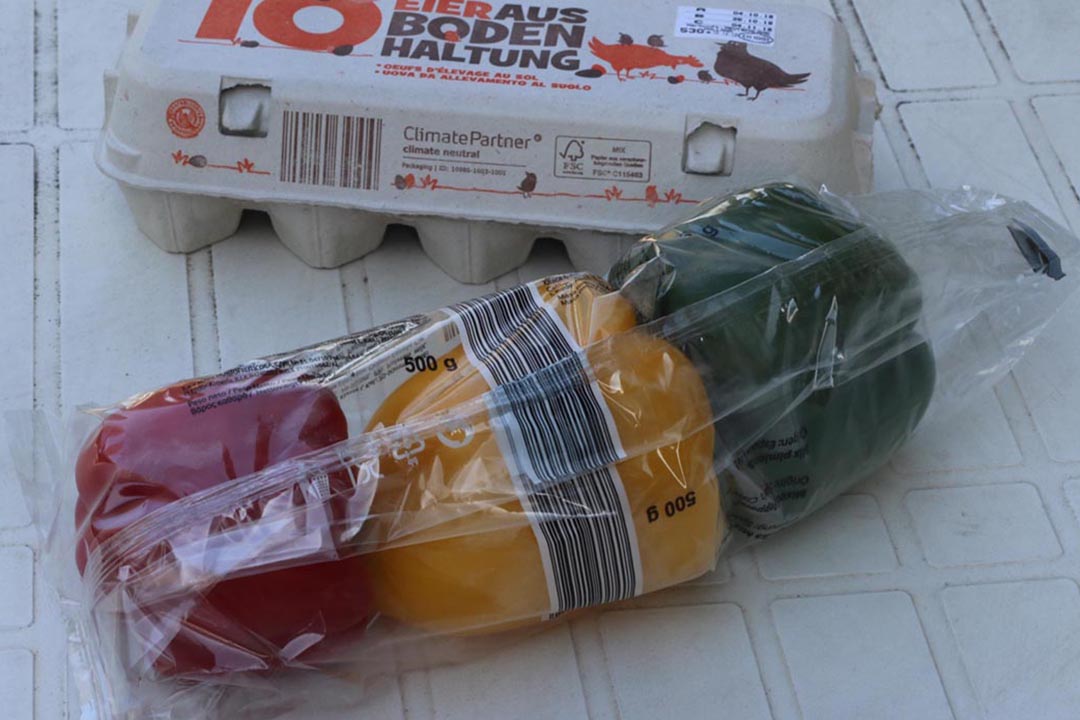}&
    \includegraphics[width=0.34\linewidth]{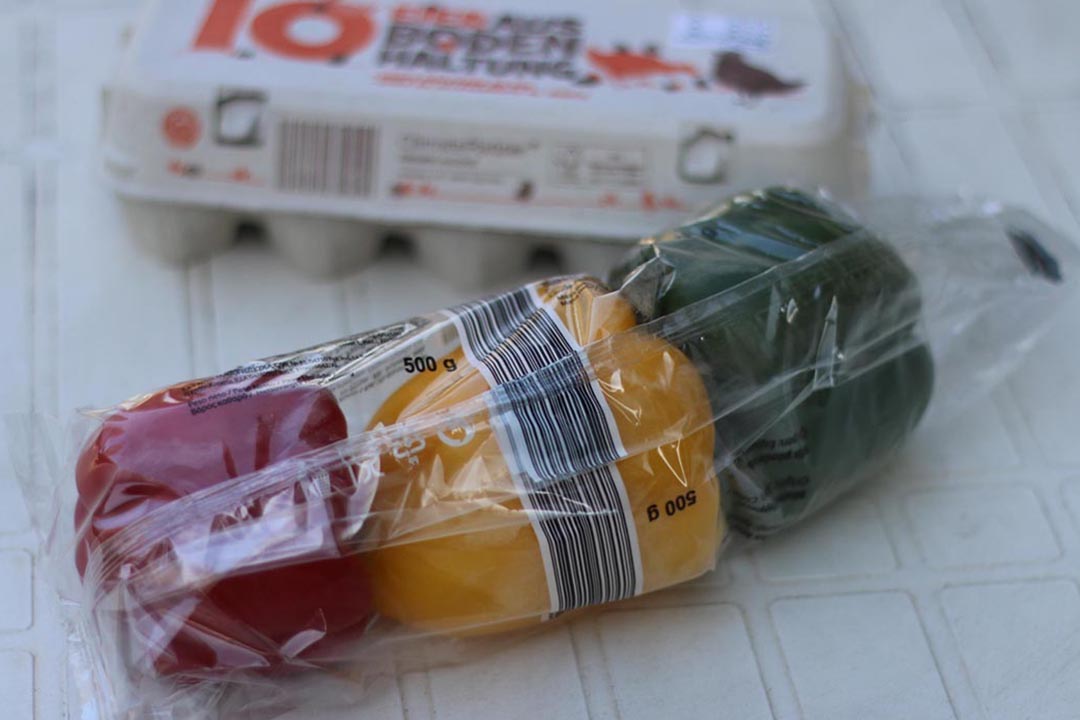} \\
    \includegraphics[width=0.34\linewidth]{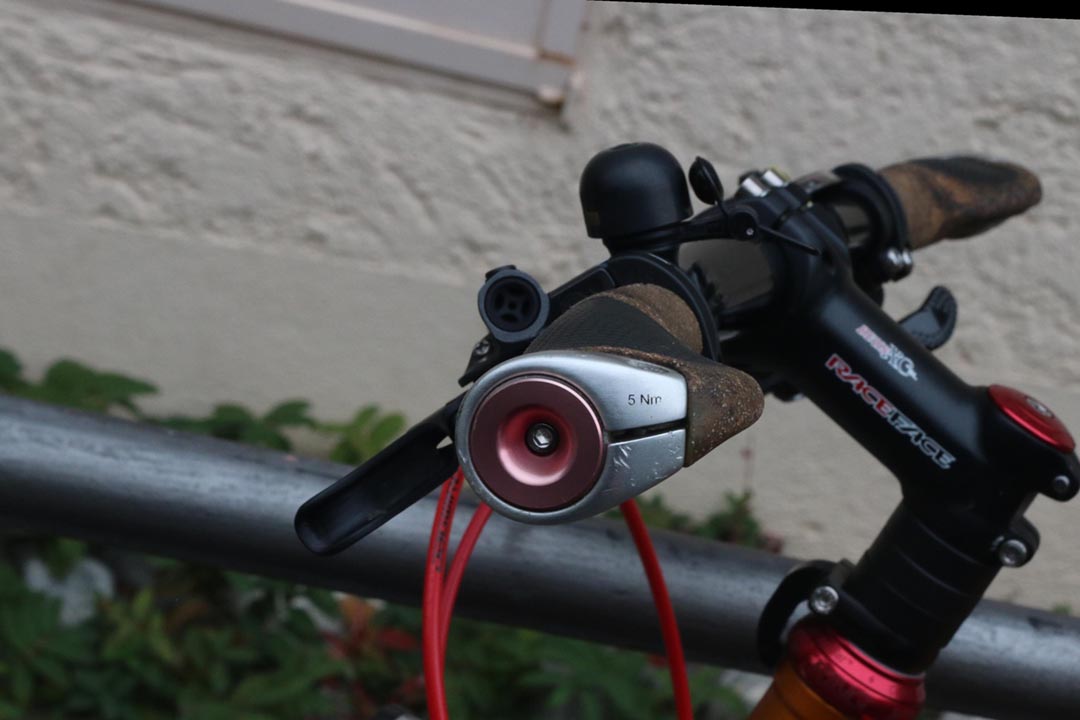}&
    \includegraphics[width=0.34\linewidth]{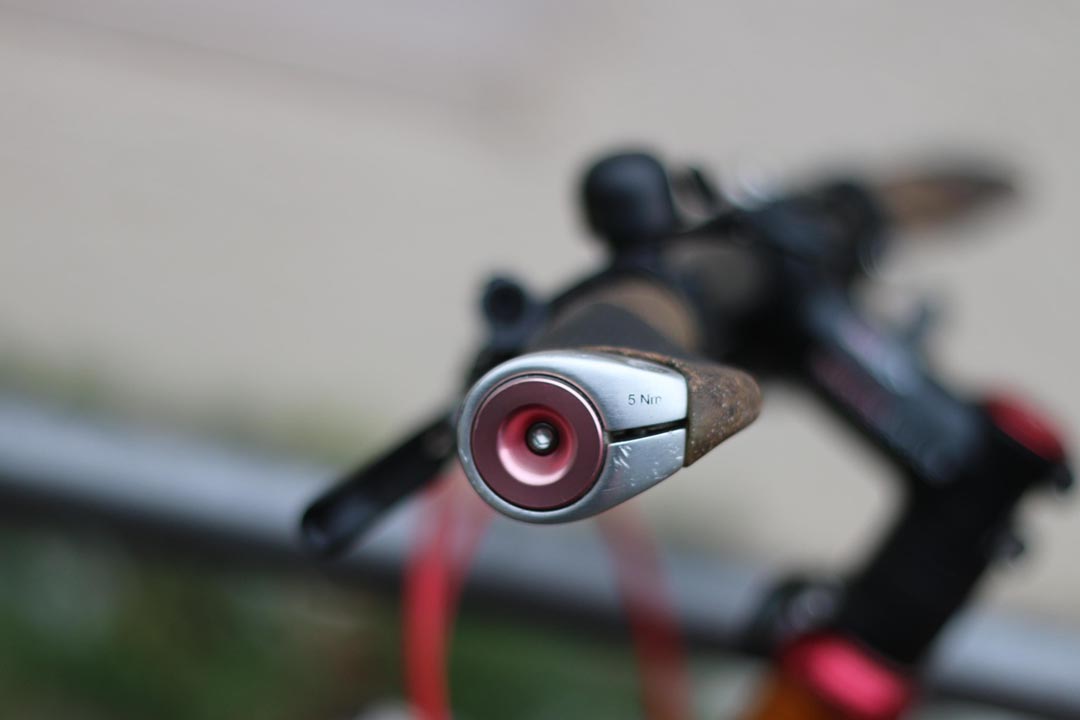}& \hspace{0.8mm}
    \includegraphics[width=0.34\linewidth]{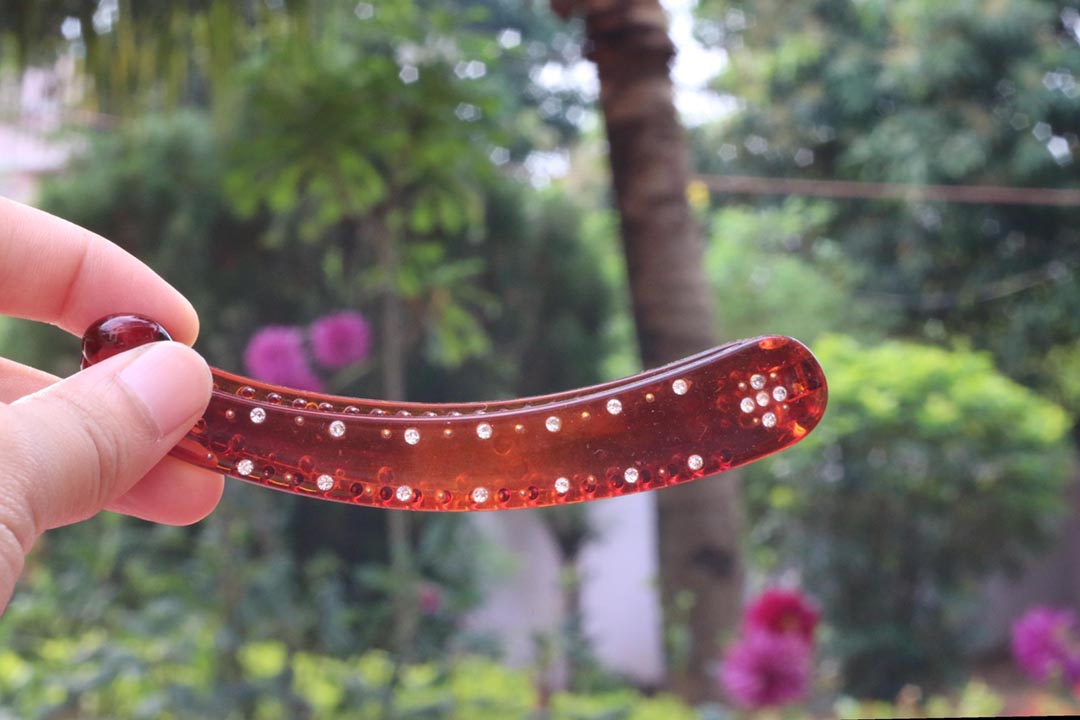}&
    \includegraphics[width=0.34\linewidth]{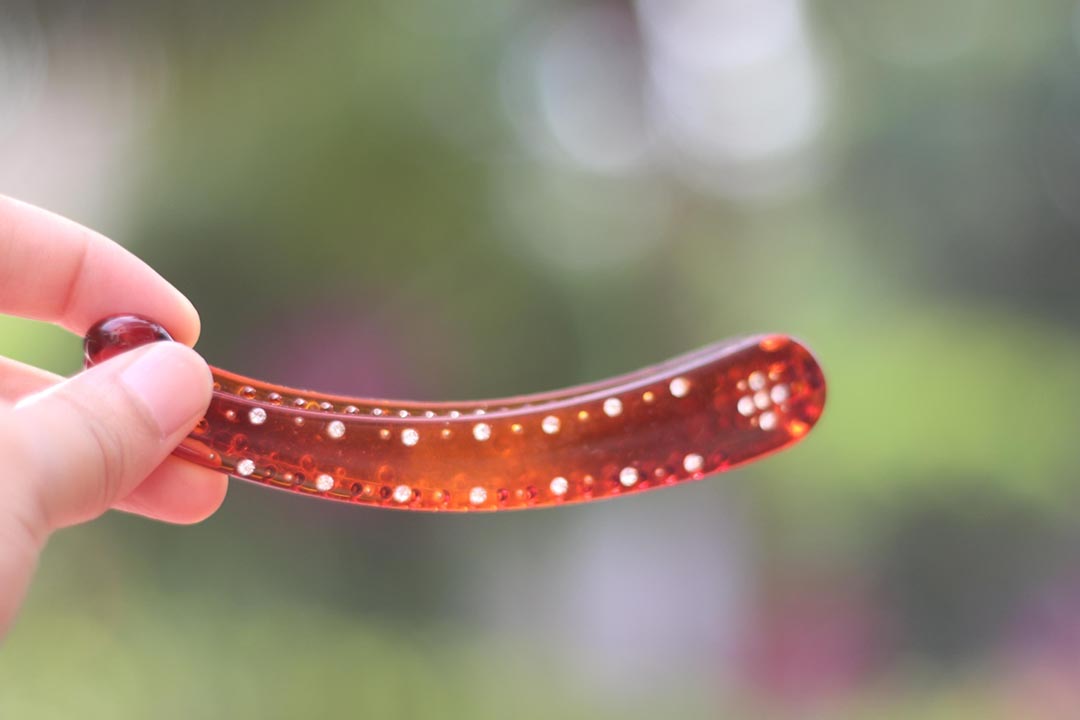}& \hspace{0.8mm}
    \includegraphics[width=0.34\linewidth]{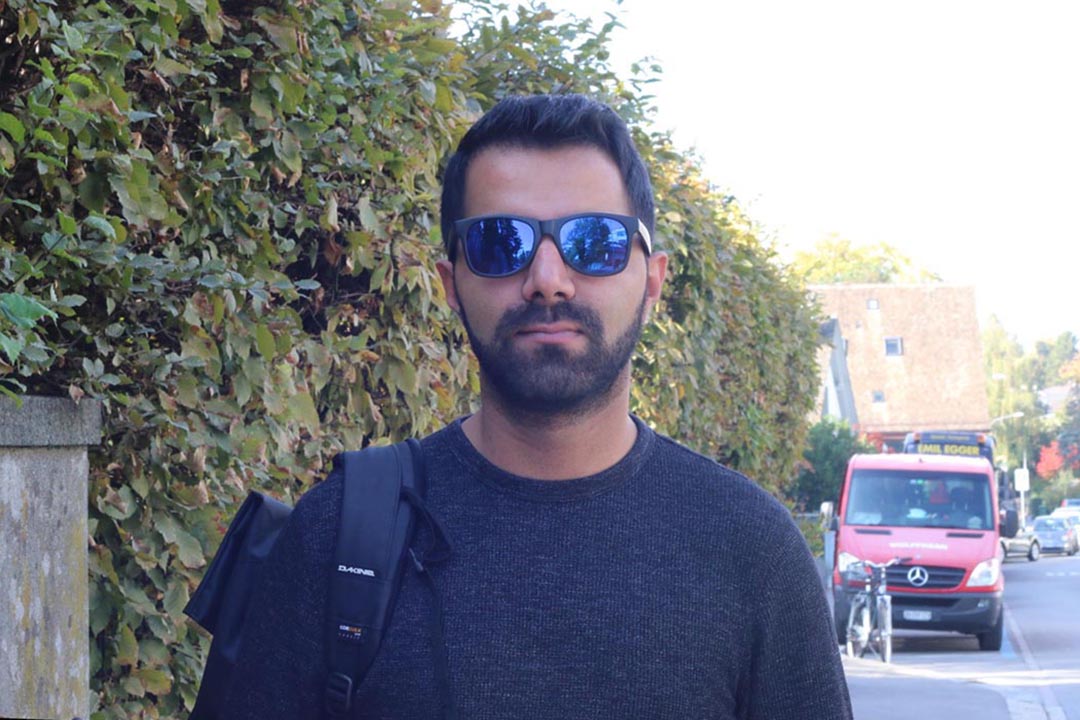}&
    \includegraphics[width=0.34\linewidth]{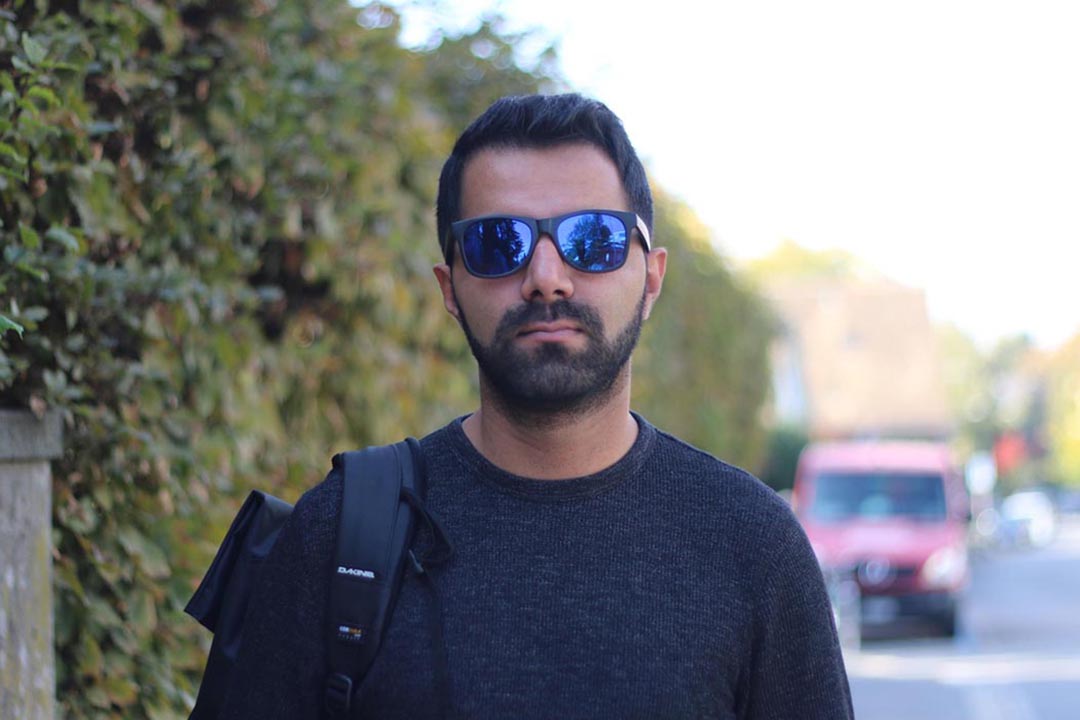} \\
    \includegraphics[width=0.34\linewidth]{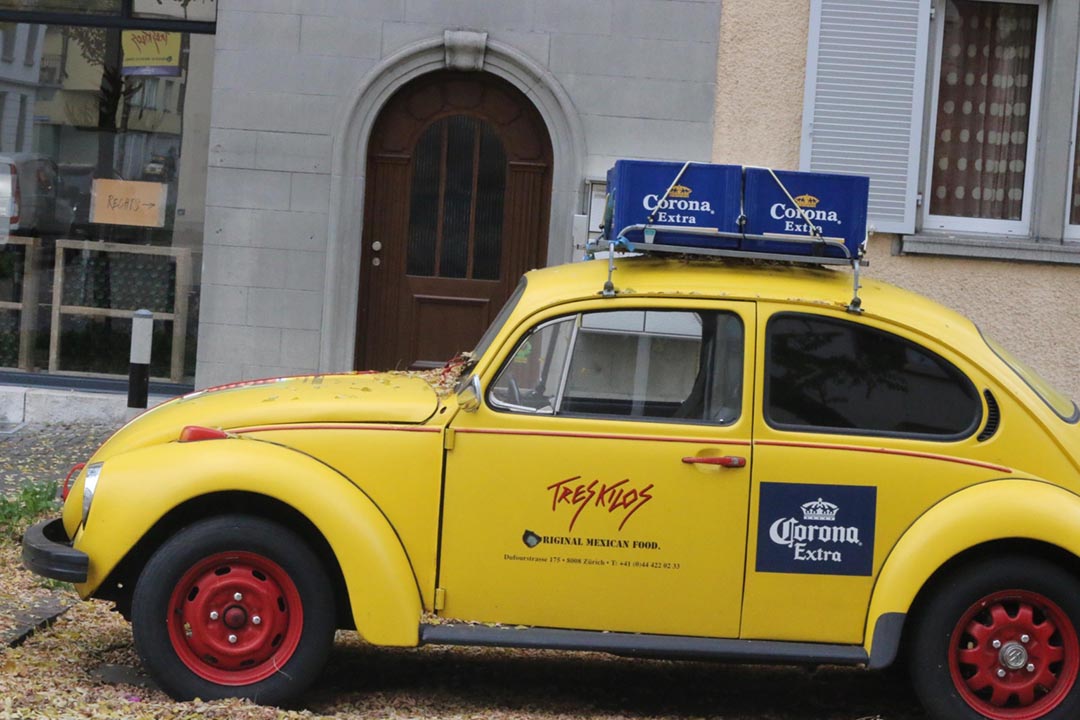}&
    \includegraphics[width=0.34\linewidth]{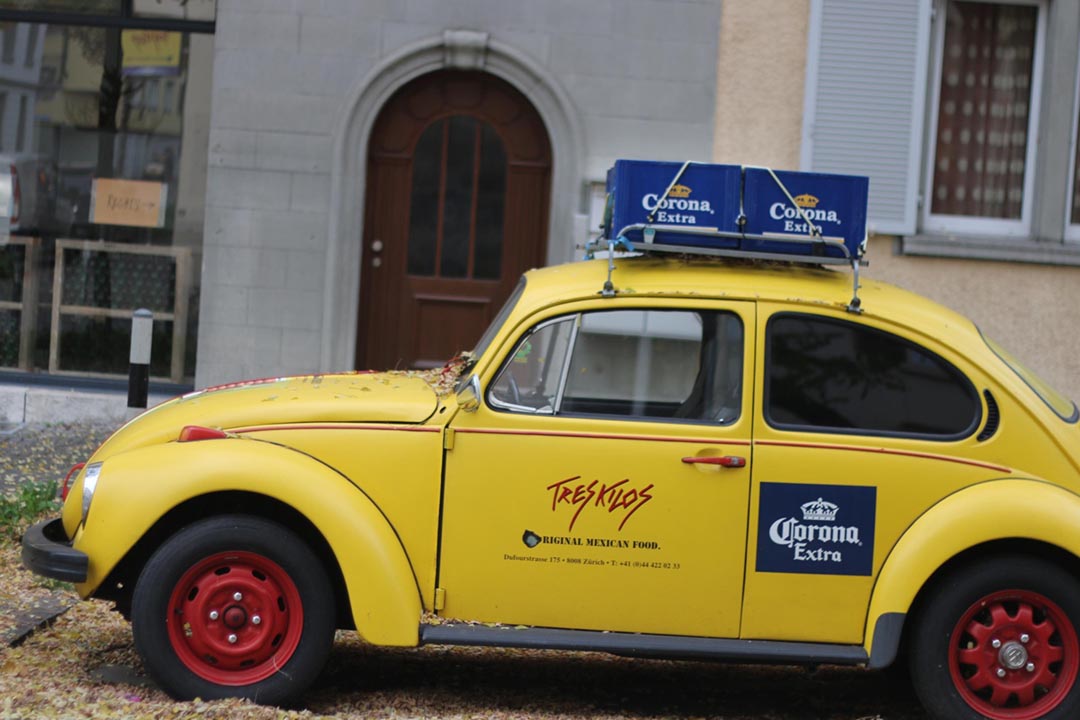}& \hspace{0.8mm}
    \includegraphics[width=0.34\linewidth]{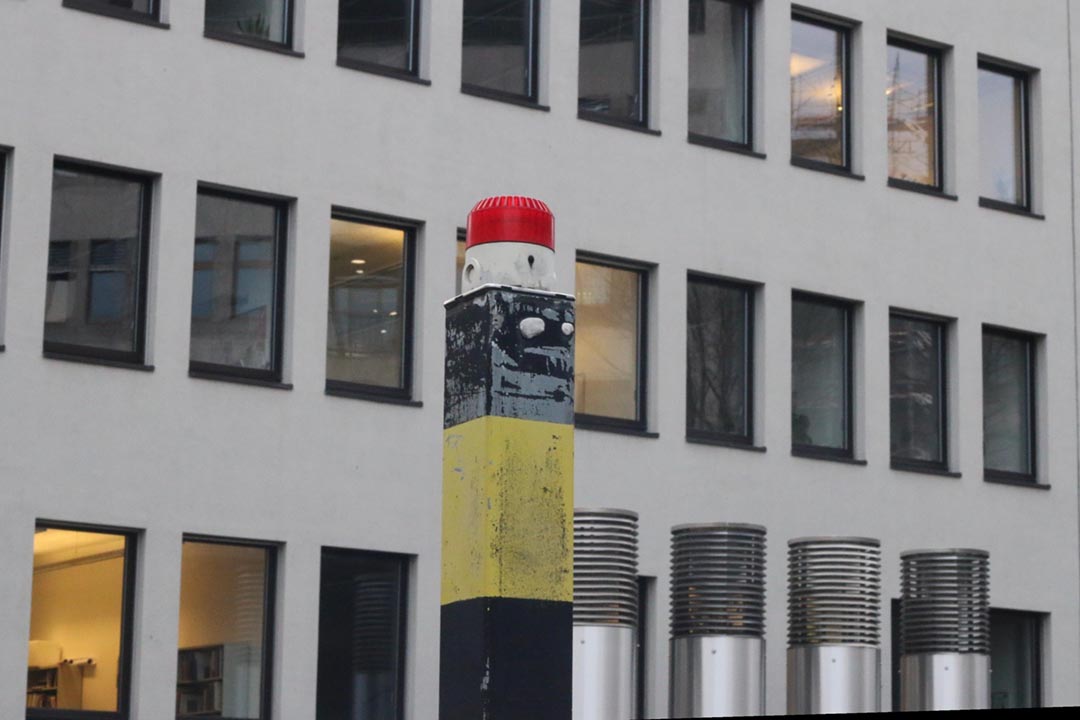}&
    \includegraphics[width=0.34\linewidth]{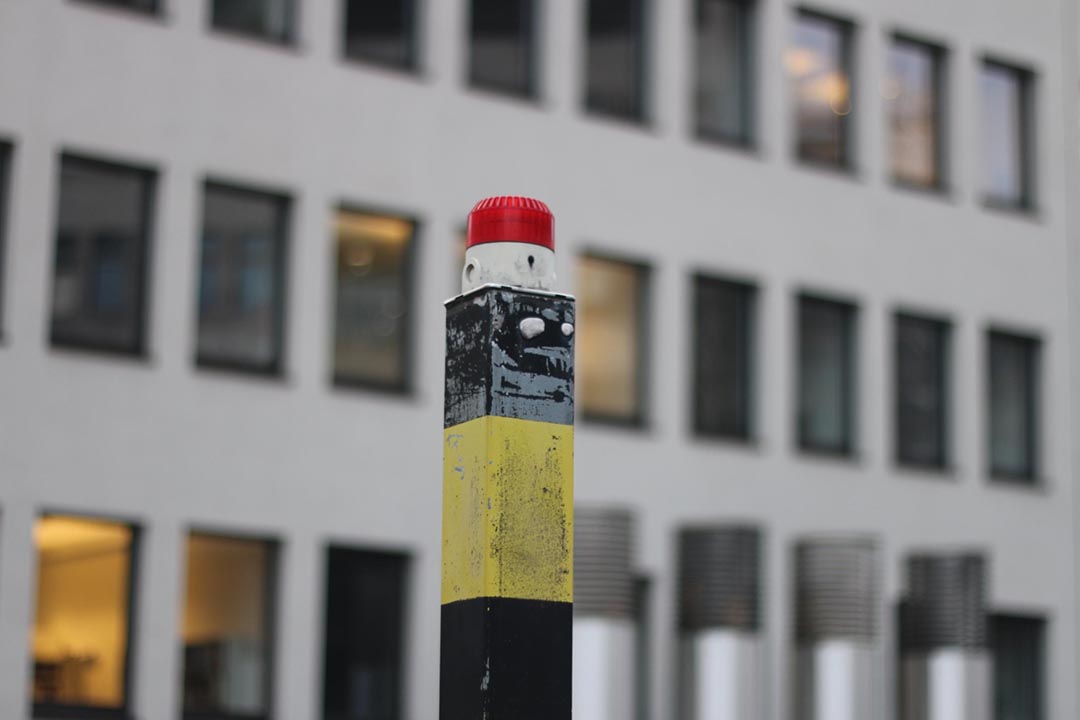}& \hspace{0.8mm}
    \includegraphics[width=0.34\linewidth]{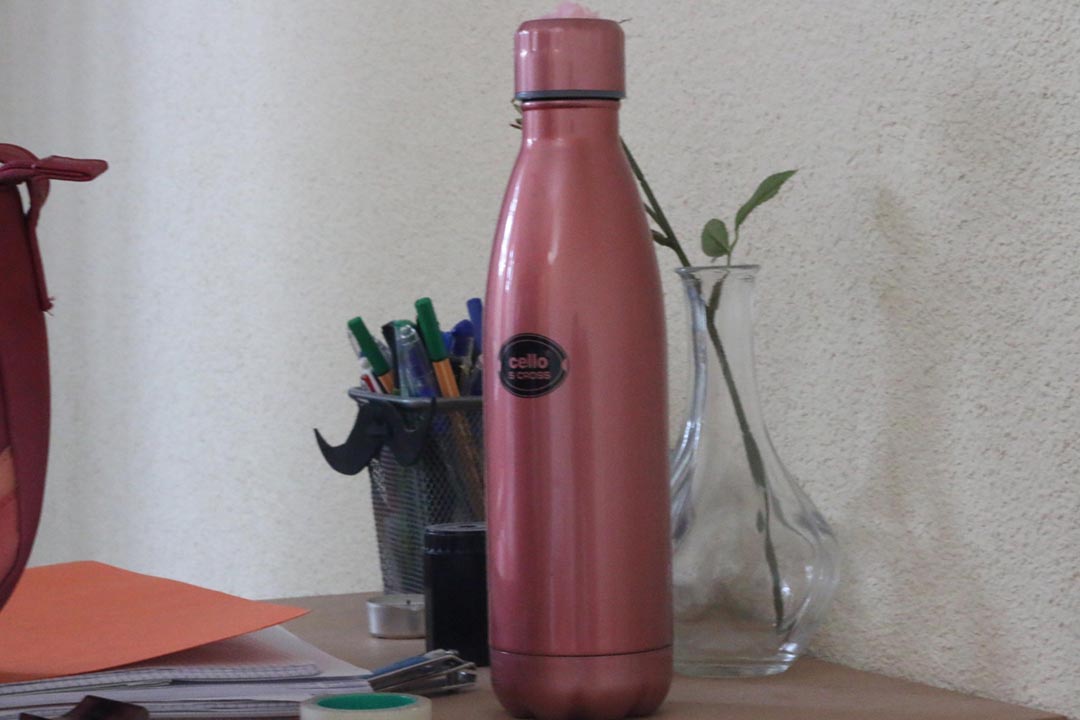}&
    \includegraphics[width=0.34\linewidth]{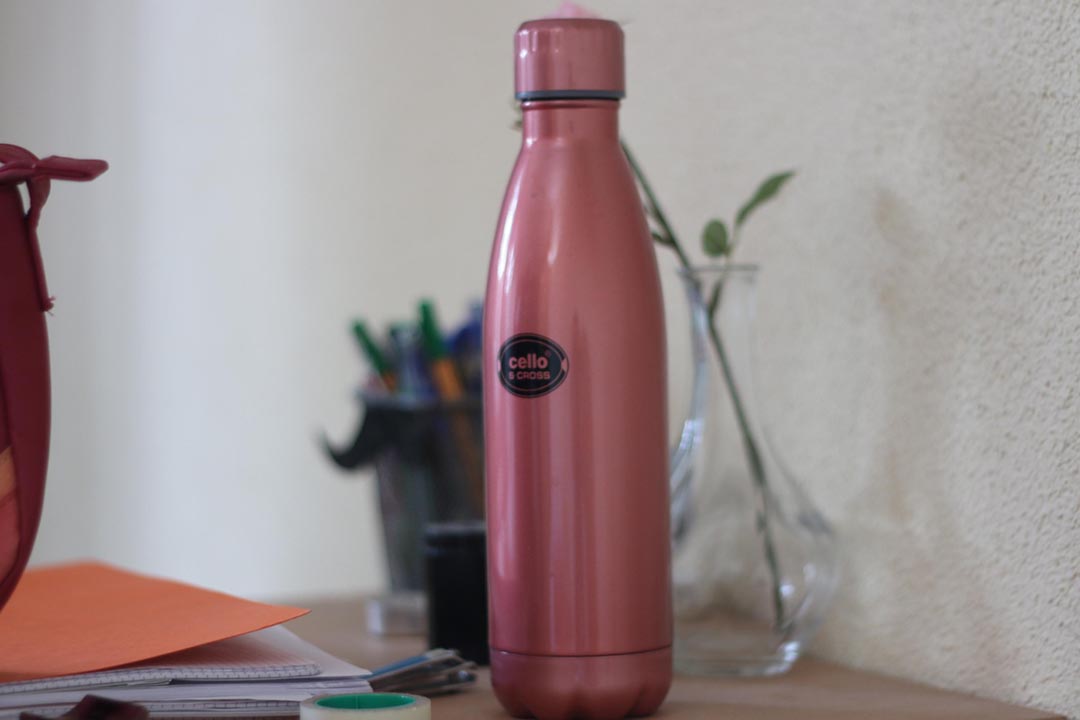} \\
    \includegraphics[width=0.34\linewidth]{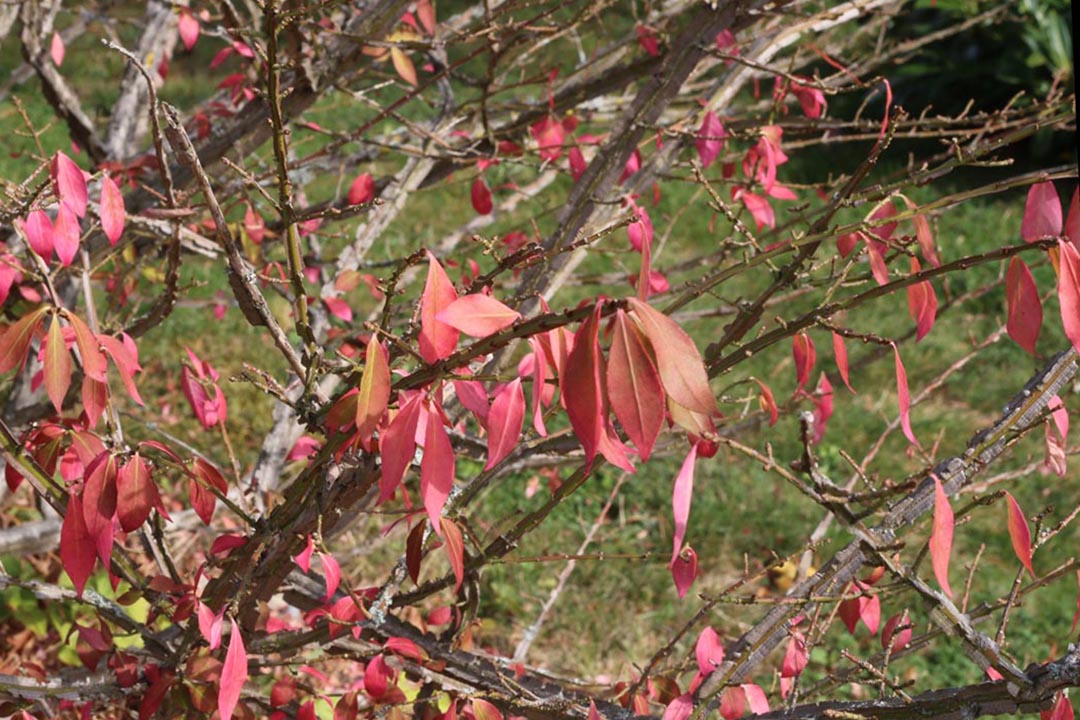}&
    \includegraphics[width=0.34\linewidth]{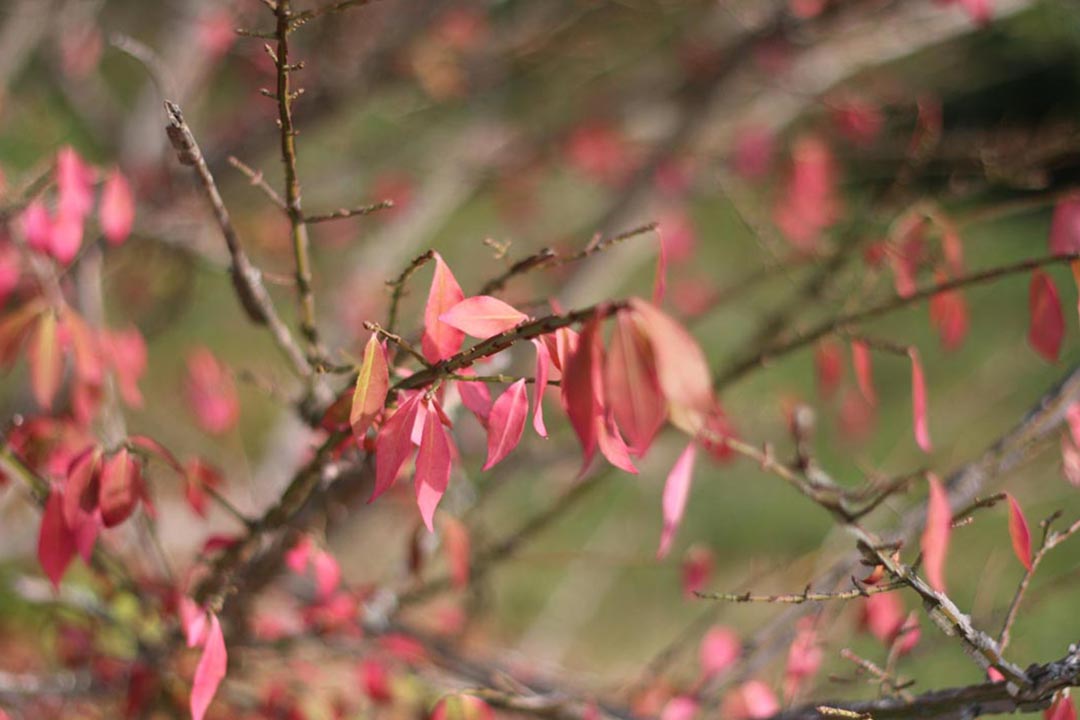}& \hspace{0.8mm}
    \includegraphics[width=0.34\linewidth]{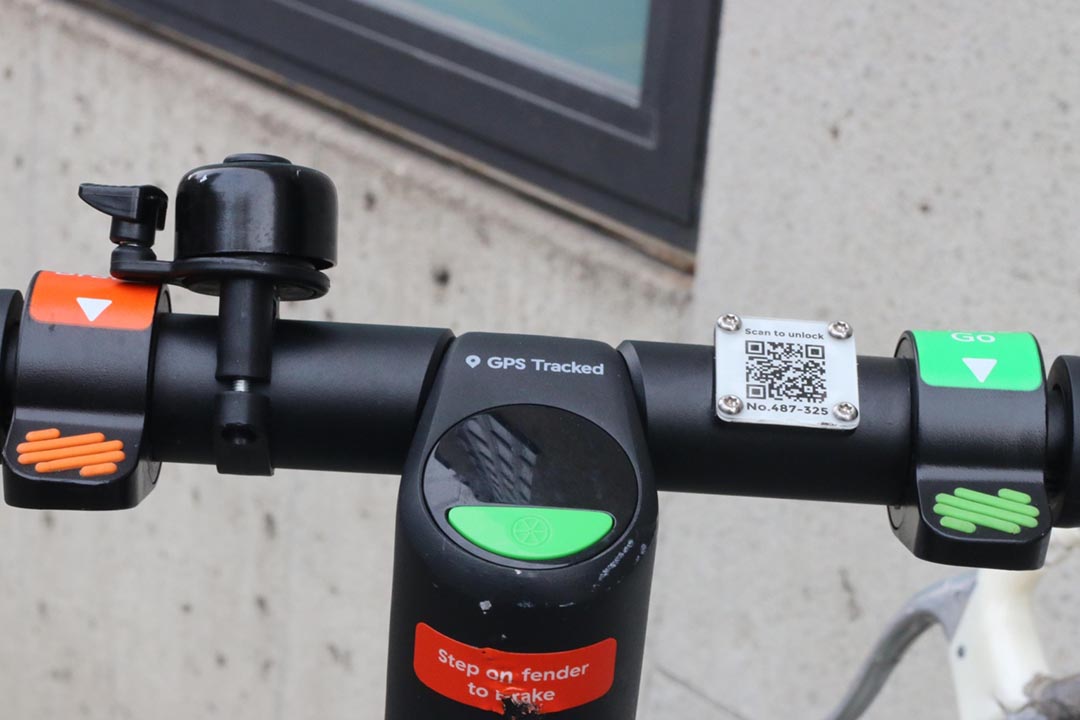}&
    \includegraphics[width=0.34\linewidth]{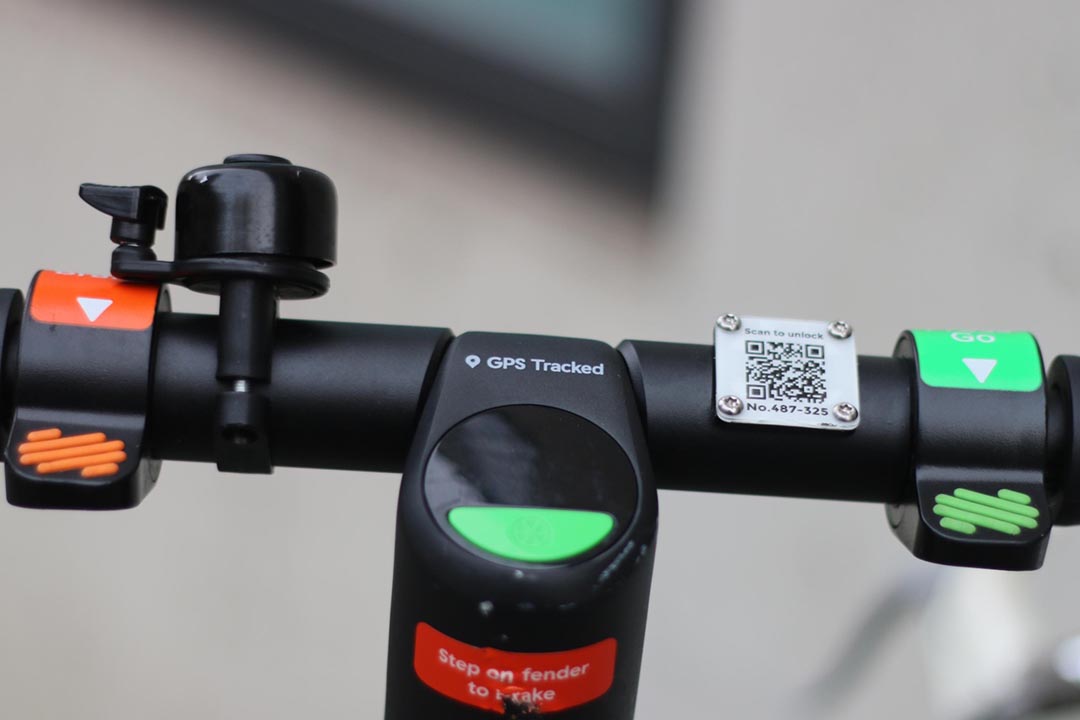}& \hspace{0.8mm}
    \includegraphics[width=0.34\linewidth]{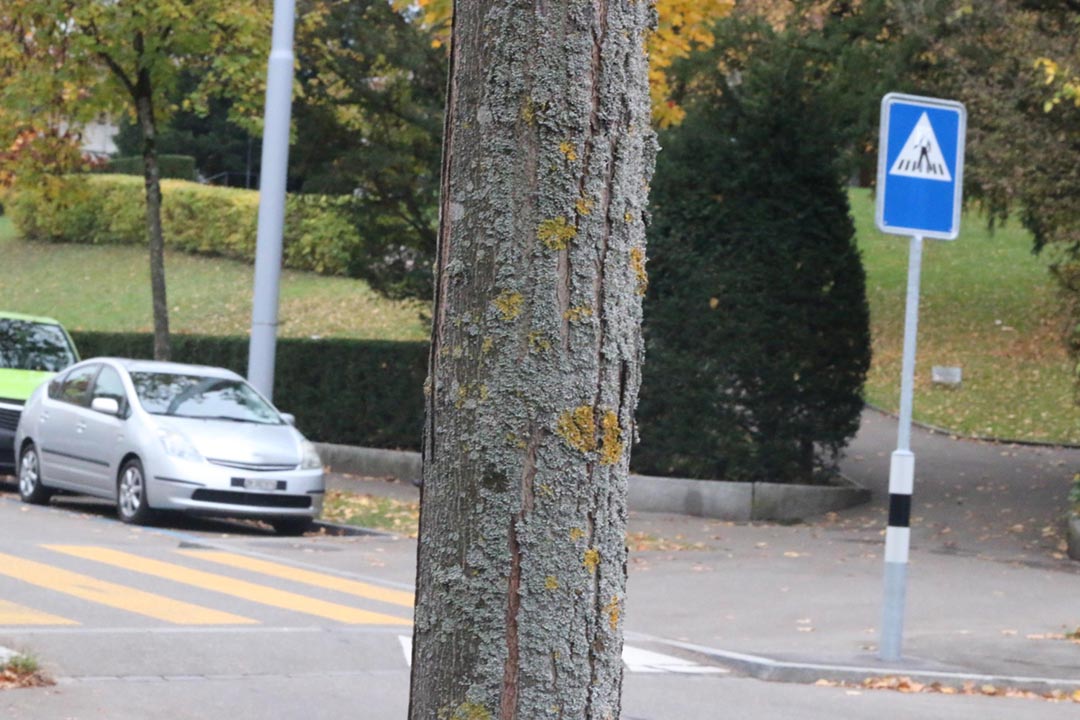}&
    \includegraphics[width=0.34\linewidth]{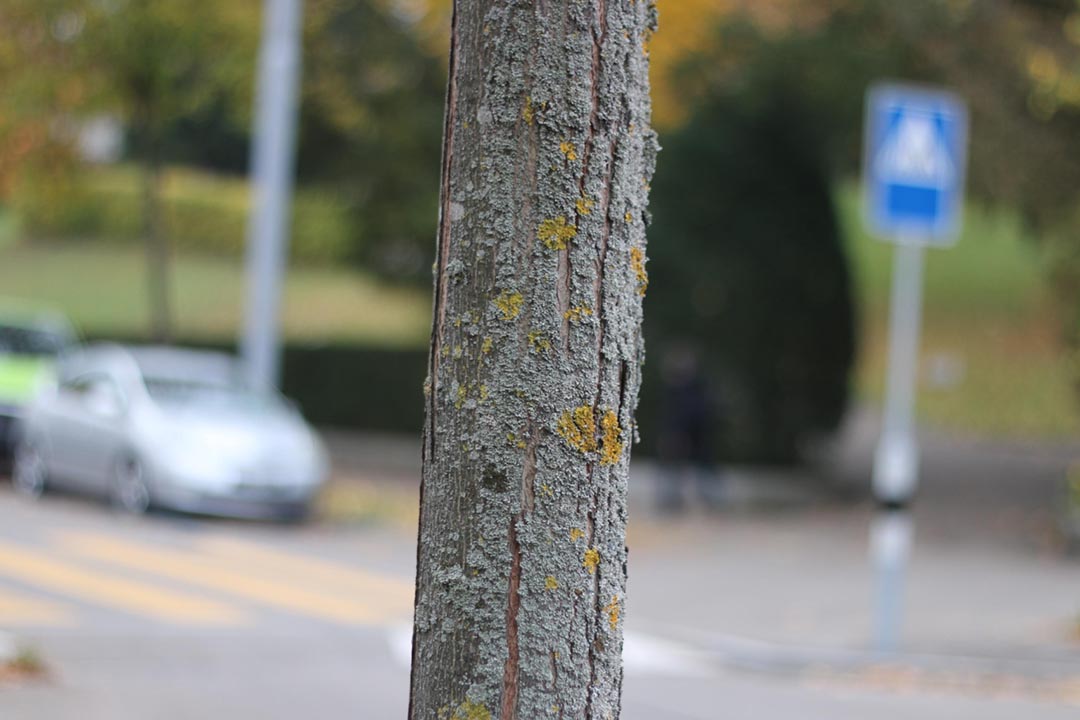} \\
    \includegraphics[width=0.34\linewidth]{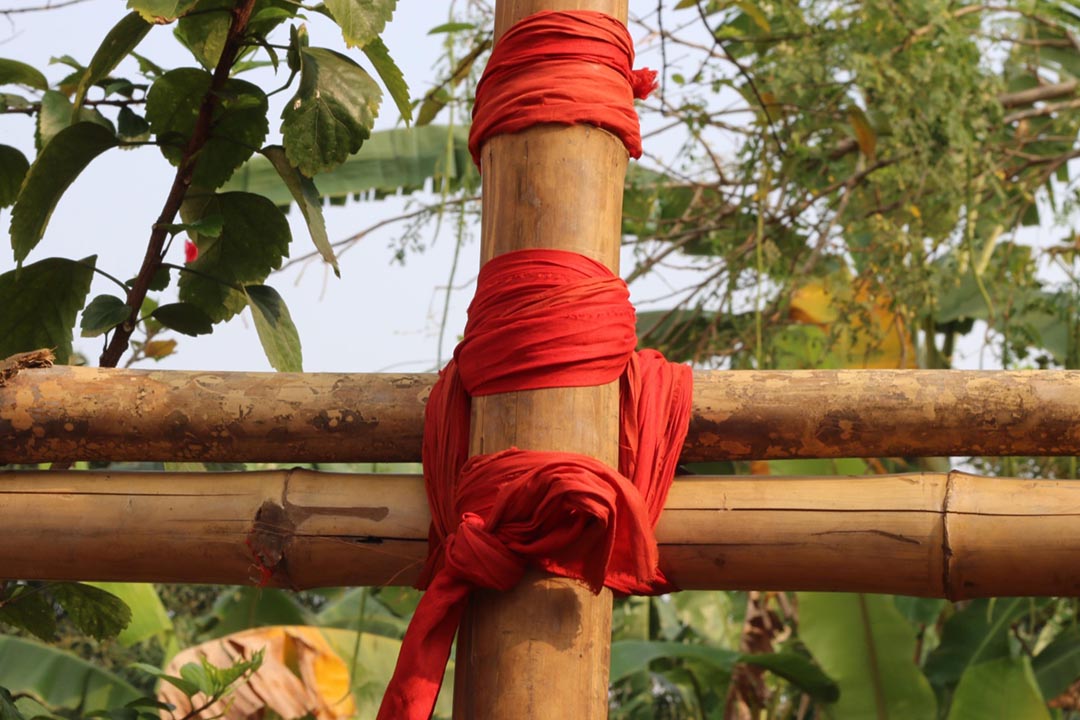}&
    \includegraphics[width=0.34\linewidth]{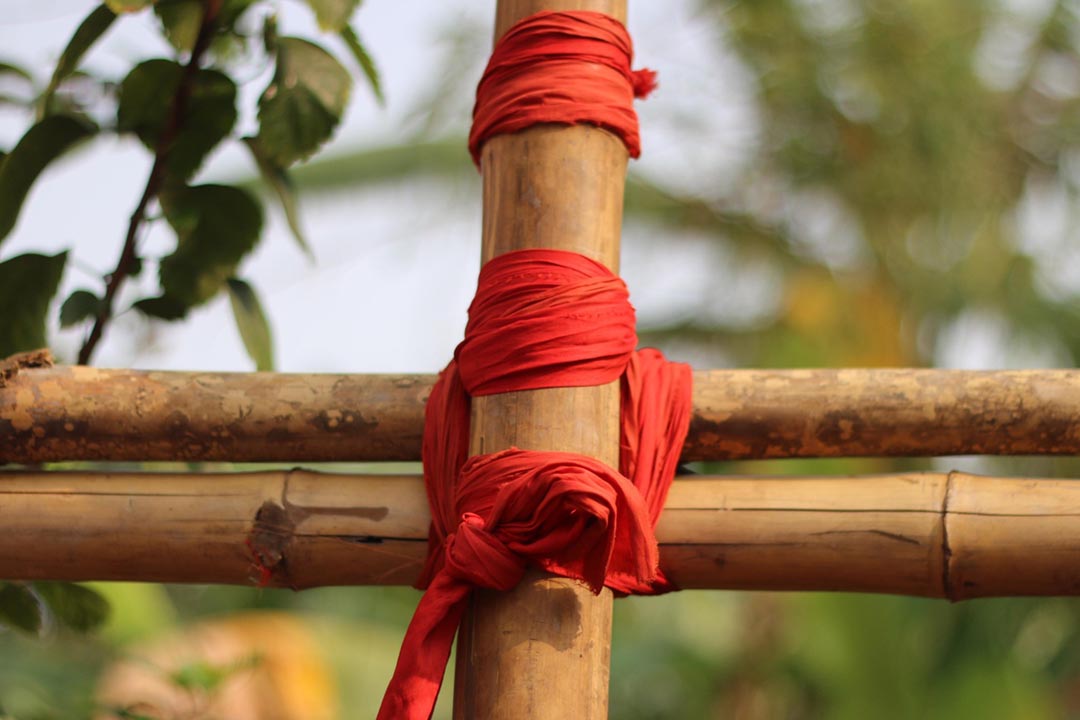}& \hspace{0.8mm}
    \includegraphics[width=0.34\linewidth]{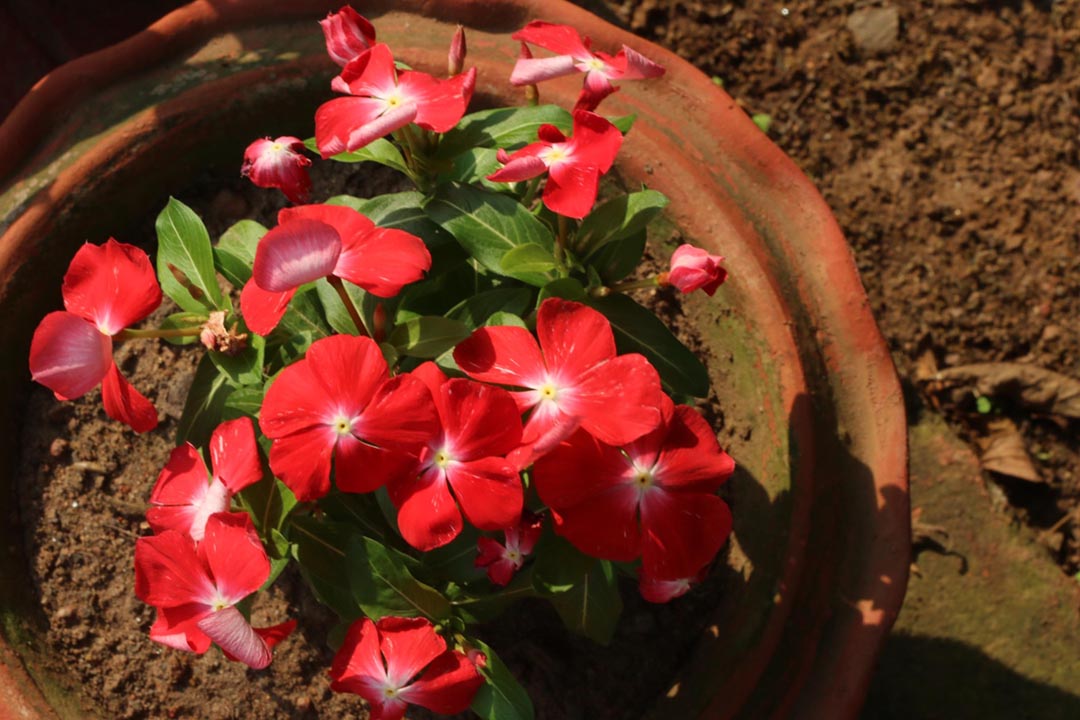}&
    \includegraphics[width=0.34\linewidth]{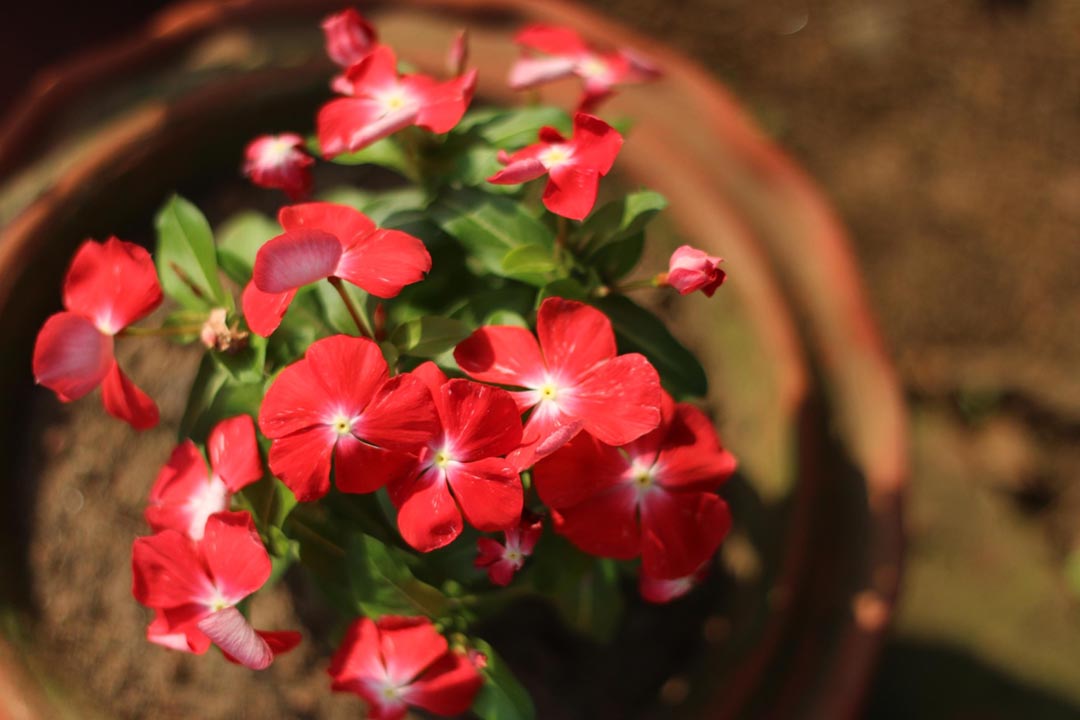}& \hspace{0.8mm}
    \includegraphics[width=0.34\linewidth]{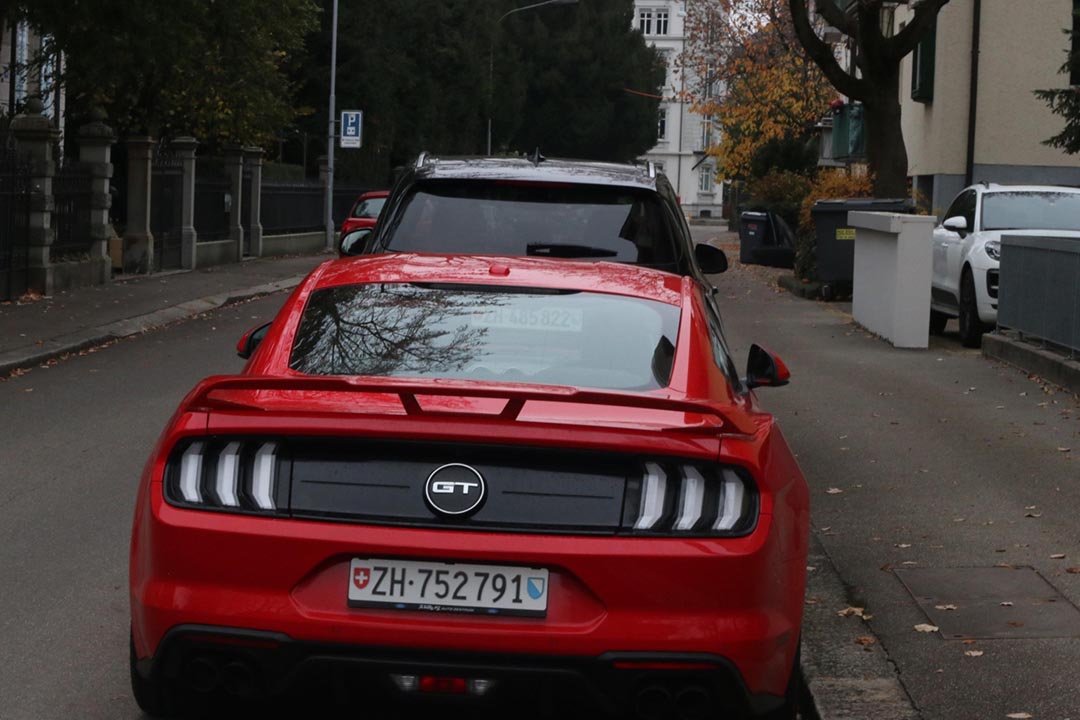}&
    \includegraphics[width=0.34\linewidth]{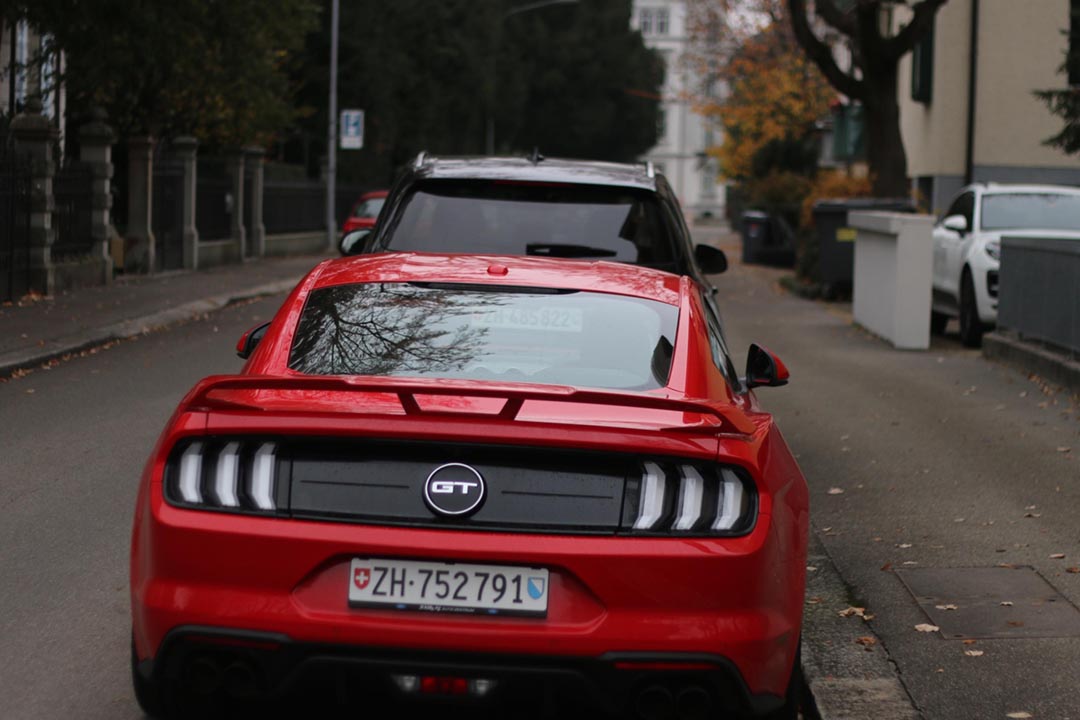} \\
\end{tabular}
}
\vspace{2.0mm}
\caption{Sample wide and shallow depth-of-field image pairs from the EBB! dataset.}
\label{fig:example_photos}
\end{figure*}

Bokeh is common technique used to highlight the main object on the photo by blurring all out-of-focus regions. Due to physical limitations of mobile cameras, they cannot produce a strong bokeh effect naturally, thus computer vision-based approaches are often used for this on mobile devices. This topic became very popular over the past few years, and one of the most common solution for this problem consists in segmenting out the main object of interest on the photo and then blurring the background~\cite{shen2016automatic,shen2016deep,zhu2017fast}. Another approach proposed for this task is to  blur the image based on the predicted depth map~\cite{Hernandez2014Lens,ha2016high,yu20143d} that can be obtained either using the parallax effect or with stereo vision~\cite{barron2015fast,fausto2022depth}. Finally, an end-to-end deep learning-based solution and the corresponding \textit{EBB!} bokeh effect rendering dataset was proposed in~\cite{ignatov2020rendering}, where the authors developed a neural network capable of transforming wide to shallow depth-of-field images automatically. Many other alternative solutions using this dataset were presented in~\cite{ignatov2019aimBokeh,ignatov2020aimBokeh,qian2020bggan,dutta2021depth,dutta2021stacked,wang2022rendering,wang2022rendering,zheng2022constrained,peng2022bokehme}.

The deployment of AI-based solutions on portable devices usually requires an efficient model design based on a good understanding of the mobile processing units (\eg CPUs, NPUs, GPUs, DSP) and their hardware particularities, including their memory constraints. We refer to~\cite{ignatov2019ai,ignatov2018ai} for an extensive overview of mobile AI acceleration hardware, its particularities and performance. As shown in these works, the latest generations of mobile NPUs are reaching the performance of older-generation mid-range desktop GPUs. Nevertheless, a straightforward deployment of neural networks-based solutions on mobile devices is impeded by (i) a limited memory (\ie, restricted amount of RAM) and
(ii) a limited or lacking support of many common deep learning operators and layers. These impeding factors make the processing of high resolution inputs impossible with the standard NN models and require a careful adaptation or re-design to the constraints of mobile AI hardware. Such optimizations can employ a combination of various model techniques such as 16-bit / 8-bit~\cite{chiang2020deploying,jain2019trained,jacob2018quantization,yang2019quantization} and low-bit~\cite{cai2020zeroq,uhlich2019mixed,ignatov2020controlling,liu2018bi} quantization, network pruning and compression~\cite{chiang2020deploying,ignatov2020rendering,li2019learning,liu2019metapruning,obukhov2020t}, device- or NPU-specific adaptations, platform-aware neural architecture search~\cite{howard2019searching,tan2019mnasnet,wu2019fbnet,wan2020fbnetv2}, \etc.

The majority of competitions aimed at efficient deep learning models use standard desktop hardware for evaluating the solutions, thus the obtained models rarely show acceptable results when running on real mobile hardware with many specific constraints. In this \textit{Mobile AI challenge}, we take a radically different approach and propose the participants to develop and evaluate their models directly on mobile devices. The goal of this competition is to design a efficient deep learning-based solution for the realistic bokeh effect rendering problem. For this, the participants were provided with a large-scale \textit{EBB!} dataset containing 5K shallow / wide depth-of-field image pairs collected in the wild with the Canon 7D DSLR camera and 50mm f/1.8 fast lens. The efficiency of the proposed solutions was evaluated on the Kirin 9000 mobile platform capable of accelerating floating-point and quantized neural networks. The majority of solutions developed in this challenge are fully compatible with the TensorFlow Lite framework~\cite{TensorFlowLite2021}, thus can be efficiently executed on various Linux and Android-based IoT platforms, smartphones and edge devices.

\smallskip

This challenge is a part of the \textit{Mobile AI \& AIM 2022 Workshops and Challenges} consisting of the following competitions:

\small

\begin{itemize}
\item Realistic Bokeh Effect Rendering on Mobile GPUs
\item Learned Smartphone ISP on Mobile GPUs~\cite{ignatov2022maiisp}
\item Power Efficient Video Super-Resolution on Mobile NPUs~\cite{ignatov2022maivideosr}
\item Quantized Image Super-Resolution on Mobile NPUs~\cite{ignatov2022maisuperres}
\item Efficient Single-Image Depth Estimation on Mobile Devices~\cite{ignatov2022maidepth}
\item Super-Resolution of Compressed Image and Video~\cite{yang2022aim}
\item Reversed Image Signal Processing and RAW Reconstruction~\cite{conde2022aim}
\item Instagram Filter Removal~\cite{kinli2022aim}
\end{itemize}

\noindent The results and solutions obtained in the previous \textit{MAI 2021 Challenges} are described in our last year papers:

\small

\begin{itemize}
\item Learned Smartphone ISP on Mobile NPUs~\cite{ignatov2021learned}
\item Real Image Denoising on Mobile GPUs~\cite{ignatov2021fastDenoising}
\item Quantized Image Super-Resolution on Mobile NPUs~\cite{ignatov2021real}
\item Real-Time Video Super-Resolution on Mobile GPUs~\cite{romero2021real}
\item Single-Image Depth Estimation on Mobile Devices~\cite{ignatov2021fastDepth}
\item Quantized Camera Scene Detection on Smartphones~\cite{ignatov2021fastSceneDetection}
\end{itemize}

\normalsize


\begin{figure*}[t!]
\centering
\setlength{\tabcolsep}{1pt}
\resizebox{0.96\linewidth}{!}
{
\includegraphics[width=1.0\linewidth]{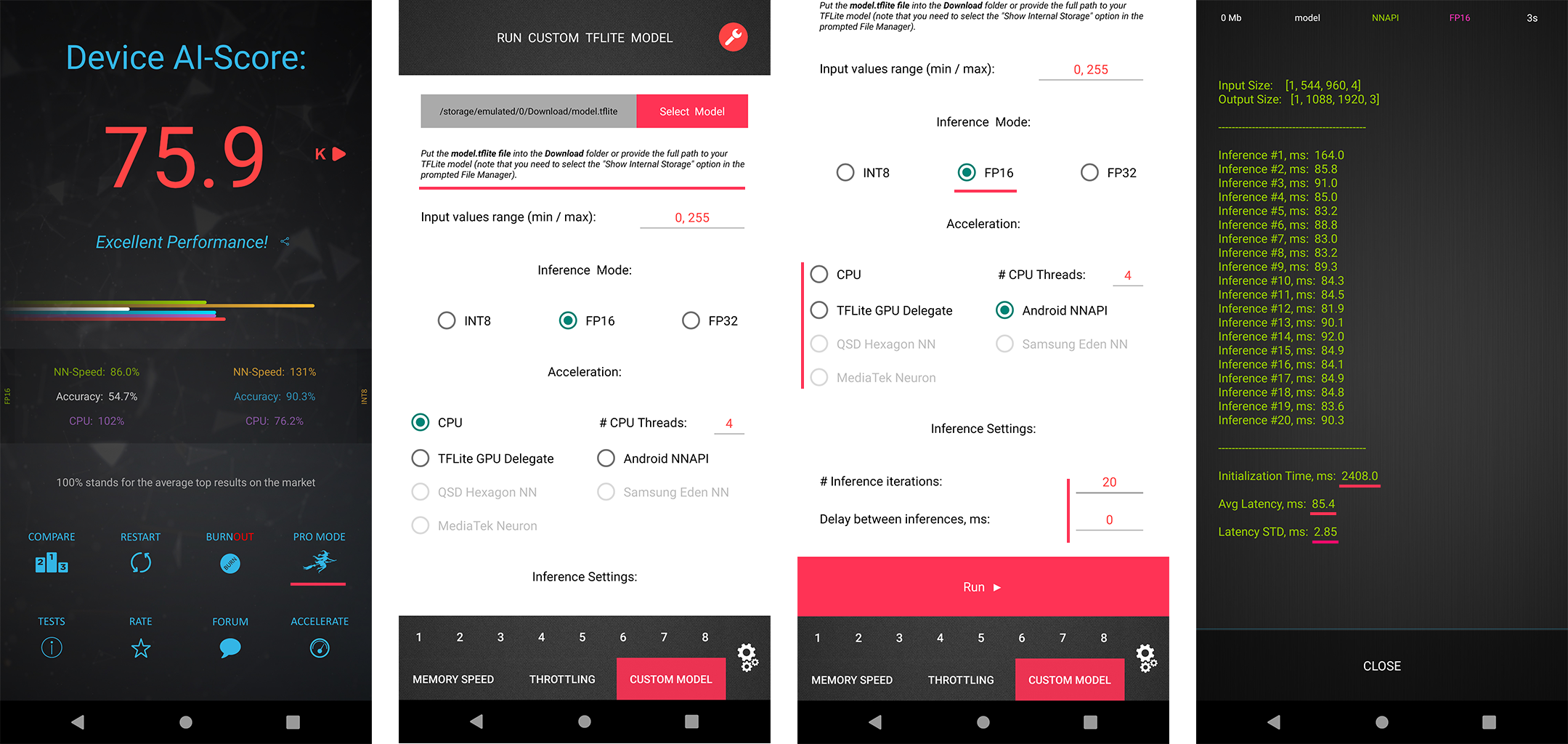}
}
\vspace{0.2cm}
\caption{Loading and running custom TensorFlow Lite models with AI Benchmark application. The currently supported acceleration options include Android NNAPI, TFLite GPU, Hexagon NN, Qualcomm QNN, MediaTek Neuron and Samsung ENN delegates as well as CPU inference through TFLite or XNNPACK backends. The latest app version can be downloaded at \url{https://ai-benchmark.com/download}.}
\label{fig:ai_benchmark_custom}
\end{figure*}

\section{Challenge}

In order to design an efficient and practical deep learning-based solution for the considered task that runs fast on mobile devices, one needs the following tools:

\begin{enumerate}
\item A large-scale high-quality dataset for training and evaluating the models. Real, not synthetically generated data should be used to ensure a high quality of the obtained model;
\item An efficient way to check the runtime and debug the model locally without any constraints as well as the ability to check the runtime on the target evaluation platform.
\end{enumerate}

This challenge addresses all the above issues. Real training data, tools, and runtime evaluation options provided to the challenge participants are described in the next sections.

\subsection{Dataset}

One of the biggest challenges in the bokeh rendering task is to get high-quality real data that can be used for training deep models. To tackle this problem, in this challenge we use a popular large-scale \textit{Everything is Better with Bokeh!} (EBB!) dataset containing more than 10 thousand images was collected in the wild during several months. By controlling the aperture size of the lens, images with shallow and wide depth-of-field were taken. In each photo pair, the first image was captured with a narrow aperture (f/16) that results in a normal sharp photo, whereas the second one was shot using the highest aperture (f/1.8) leading to a strong bokeh effect. The photos were taken during the daytime in a wide variety of places and in various illumination and weather conditions. The photos were captured in automatic mode, the default settings were used throughout the entire collection procedure. An example set of collected images is presented in Figure~\ref{fig:example_photos}.

The captured image pairs are not aligned exactly, therefore they were first matched using SIFT keypoints and RANSAC method same as in~\cite{ignatov2017dslr}. The resulting images were then cropped to their intersection part and downscaled so that their final height is equal to 1024 pixels. From the resulting 10 thousand images, 200 image pairs were reserved for testing, while the other 4.8 thousand photo pairs can be used for training and validation.

\subsection{Local Runtime Evaluation}

When developing AI solutions for mobile devices, it is vital to be able to test the designed models and debug all emerging issues locally on available devices. For this, the participants were provided with the \textit{AI Benchmark} application~\cite{ignatov2018ai,ignatov2019ai} that allows to load any custom TensorFlow Lite model and run it on any Android device with all supported acceleration options. This tool contains the latest versions of \textit{Android NNAPI, TFLite GPU, Hexagon NN, Qualcomm QNN, MediaTek Neuron} and \textit{Samsung ENN} delegates, therefore supporting all current mobile platforms and providing the users with the ability to execute neural networks on smartphone NPUs, APUs, DSPs, GPUs and CPUs.

\smallskip

To load and run a custom TensorFlow Lite model, one needs to follow the next steps:

\begin{enumerate}
\setlength\itemsep{0mm}
\item Download AI Benchmark from the official website\footnote{\url{https://ai-benchmark.com/download}} or from the Google Play\footnote{\url{https://play.google.com/store/apps/details?id=org.benchmark.demo}} and run its standard tests.
\item After the end of the tests, enter the \textit{PRO Mode} and select the \textit{Custom Model} tab there.
\item Rename the exported TFLite model to \textit{model.tflite} and put it into the \textit{Download} folder of the device.
\item Select mode type \textit{(INT8, FP16, or FP32)}, the desired acceleration/inference options and run the model.
\end{enumerate}

\noindent These steps are also illustrated in Fig.~\ref{fig:ai_benchmark_custom}.

\subsection{Runtime Evaluation on the Target Platform}

In this challenge, we use the the \textit{Kirin 9000} mobile SoC as our target runtime evaluation platform. The considered chipset demonstrates very decent AI Benchmark scores and can be found in the Huawei Mate 40 Pro / X2 smartphone series. It can efficiently accelerate floating-point networks on its Mali-G78 MP24 GPU with a theoretical FP16 performance of 4.6 TFLOPS. The models were parsed and accelerated using the TensorFlow Lite GPU delegate~\cite{lee2019device} demonstrating the best performance on this platform when using general deep learning models. All final solutions were tested using the aforementioned AI Benchmark application.

\subsection{Challenge Phases}

\begin{table*}[t!]
\centering
\resizebox{\linewidth}{!}
{
\begin{tabular}{l|c|cc|ccc|cc|c}
\hline
Team \, & \, Author \, & \, Framework \, & \, Model Size, MB \, & \, PSNR$\uparrow$ \, & \, SSIM$\uparrow$ \, & \, LPIPS$\downarrow$ \, & \,CPU Runtime, ms $\downarrow$ \, & GPU Runtime, ms $\downarrow$ \, & \, Final Score \\
\hline
\hline
Antins\_cv & xiaokaoji & Keras / TensorFlow & 0.06 & 22.76 & 0.8652 & 0.2693 & \textBF{125} & \textBF{28.1} & 74 \\
ENERZAi & MinsuKwon & TensorFlow & 30 & 22.89 & 0.8754 & 0.2464 & 1637 & 89.3 & 28 \\
MiAIgo & hxin & TensorFlow & 1.5 & 20.08 & 0.7209 & 0.4349 & 1346 & 112 & 0.5 \\
Sensebrain & brianjsl & \, PyTorch / TensorFlow \, & 402 & 22.81 & 0.8653 & 0.2207 & 12879 & Error & N.A. \\
ZJUT-Vision \, & \, HaotianQian \, & TensorFlow & 13 & \textBF{23.53} & \textBF{0.8796} & \textBF{0.1907} & Error & Error & N.A. \\
VIC & hyessol & PyTorch & 127 & 22.77 & 0.8713 & 0.2657 & Error & Error & N.A. \\
\hline
PyNET~\cite{ignatov2020rendering} & Baseline & TensorFlow & 181 & 23.28 & 0.8780 & 0.2438 & N.A. & 3512 & 1.2 \\
\end{tabular}
}
\vspace{2.6mm}
\caption{\small{Mobile AI 2022 bokeh effect rendering challenge results and final rankings. The runtime values were obtained on 1024$\times$1024 pixel images on the Kirin 9000 mobile platform. The results of the PyNET-V2 model are provided for the reference.}}
\label{tab:results}
\end{table*}

The challenge consisted of the following phases:

\vspace{-0.8mm}
\begin{enumerate}
\item[I.] \textit{Development:} the participants get access to the data and AI Benchmark app, and are able to train the models and evaluate their runtime locally;
\item[II.] \textit{Validation:} the participants can upload their models to the remote server to check the fidelity scores on the validation dataset, and to compare their results on the validation leaderboard;
\item[III.] \textit{Testing:} the participants submit their final results, codes, TensorFlow Lite models, and factsheets.
\end{enumerate}
\vspace{-0.8mm}

\subsection{Scoring System}

All solutions were evaluated using the following metrics:

\vspace{-0.8mm}
\begin{itemize}
\setlength\itemsep{-0.2mm}
\item Peak Signal-to-Noise Ratio (PSNR) measuring fidelity score,
\item Structural Similarity Index Measure (SSIM) and LPIPS~\cite{zhang2018unreasonable} (proxy for perceptual score),
\item The runtime on the target Kirin 9000 platform.
\end{itemize}
\vspace{-0.8mm}

In this challenge, the participants were able to submit their final models to two tracks. In the first track, the score of each final submission was evaluated based on the next formula ($C$ is a constant normalization factor):

\smallskip
\begin{equation*}
\text{Final Score} \,=\, \frac{2^{2 \cdot \text{PSNR}}}{C \cdot \text{runtime}},
\end{equation*}
\smallskip

In the second track, all submissions were evaluated only based on their visual results as measured by the corresponding Mean Opinion Scores (MOS). This was done to allow the participants to develop larger and more powerful models for the considered task.

During the final challenge phase, the participants did not have access to the test dataset. Instead, they had to submit their final TensorFlow Lite models that were subsequently used by the challenge organizers to check both the runtime and the fidelity results of each submission under identical conditions. This approach solved all the issues related to model overfitting, reproducibility of the results, and consistency of the obtained runtime/accuracy values.

\begin{table*}[t!]
\centering
\resizebox{\linewidth}{!}
{
\begin{tabular}{l|c|cc|ccc|c}
\hline
Team \, & \, Author \, & \, Framework \, & \, Model Size, MB \, & \, PSNR$\uparrow$ \, & \, SSIM$\uparrow$ \, & \, LPIPS$\downarrow$ \, & \, MOS Score \, \\
\hline
\hline
Antins\_cv & xiaokaoji & Keras / TensorFlow & 0.06 & 22.76 & 0.8652 & 0.2693 & 2.6 \\
ENERZAi & MinsuKwon & TensorFlow & 30 & 22.89 & 0.8754 & 0.2464 & 3.5 \\
MiAIgo & hxin & TensorFlow & 1.5 & 20.08 & 0.7209 & 0.4349 & 2.3 \\
Sensebrain & brianjsl & \, PyTorch / TensorFlow \, & 402 & 22.81 & 0.8653 & 0.2207 & 3.4 \\
ZJUT-Vision \, & \, HaotianQian \, & TensorFlow & 13 & \textBF{23.53} & \textBF{0.8796} & \textBF{0.1907} & 3.8 \\
VIC & hyessol & PyTorch & 127 & 22.77 & 0.8713 & 0.2657 & 3.4 \\
\hline
PyNET~\cite{ignatov2020rendering} & Baseline & TensorFlow & 181 & 23.28 & 0.8780 & 0.2438 & 4.0 \\
\end{tabular}
}
\vspace{2.6mm}
\caption{\small{Mean Opinion Scores (MOS) of all solutions submitted during the final phase of the MAI 2022 bokeh effect rendering challenge. Visual results were assessed based on the reconstructed full resolution images. The results of the PyNET-V2 model are provided for the reference.}}
\label{tab:results_mos}
\end{table*}

\section{Challenge Results}

From the above 90 registered participants, 6 teams entered the final phase and submitted valid results, TFLite models, codes, executables, and factsheets. The proposed methods are described in Section~\ref{sec:solutions}, and the team members and affiliations are listed in Appendix~\ref{sec:apd:team}.
on
\subsection{Results and Discussion}

Tables~\ref{tab:results} and~\ref{tab:results_mos} demonstrate the fidelity, runtime and MOS results of all solutions submitted during the final test phase. Models submitted to the 1st and 2nd challenge tracks were evaluated together since the participants had to upload the corresponding TensorFlow Lite models in both cases. In the 1st tack, the overall best results were achieved by team \textit{Antins\_cv}. The authors proposed a small U-Net based model that demonstrated a very good efficiency on the Kirin 9000 SoC, being able to achieve over 30 FPS on its Mali GPU when processing 1024$\times$1024 pixel images. Very good runtime and fidelity results were also demonstrated by model designed by team \textit{ENERZAi}: the visual results of this solution were substantially better while its runtime was still less than 90 ms on the target platform.

After evaluating the visual results of the proposed models, we can see that conventional fidelity scores do not necessarily correlate with the perceptual image reconstruction quality. In particular, all solutions except for the baseline PyNET model and the network proposed by team \textit{ZJUT-Vision} produced images that suffered from different corruptions, \textit{e.g.}, issues with texture rendering in the blurred areas, significant resolution drop on the entire image, or almost no changes compared to the input image, which can be observed in case of model trained by team \textit{MiAIgo}. These results highlighted again the difficulty of an accurate assessment of the results obtained in bokeh effect rendering task as the standard metrics are often not indicating the real image quality.

\section{Challenge Methods}
\label{sec:solutions}

\noindent This section describes solutions submitted by all teams participating in the final stage of the MAI 2022 Bokeh Effect Rendering challenge.

\subsection{Antins\_cv}

\begin{figure}[h!]
\centering
\resizebox{1.0\linewidth}{!}
{
\includegraphics[width=1.0\linewidth]{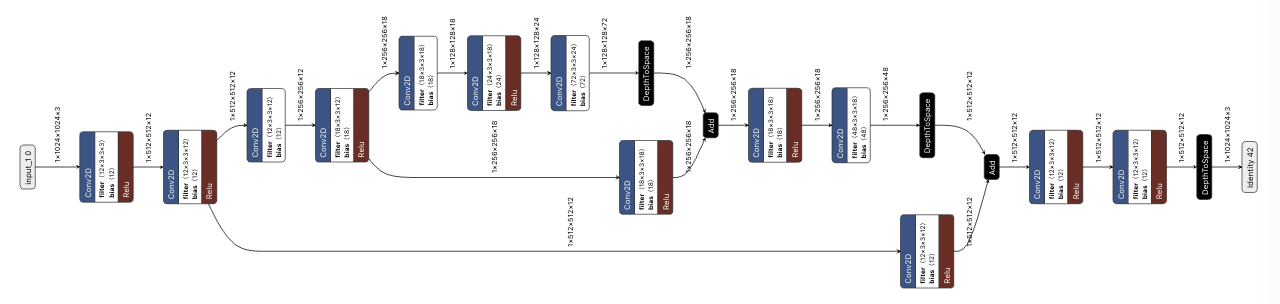}
}
\caption{\small{Model architecture proposed by team Antins\_cv.}}
\label{fig:Antins}
\end{figure}

Team Antins\_cv proposed a tiny U-Net based model for the considered task, which architecture is illustrated in Fig.~\ref{fig:Antins}. The model is composed of three levels: the first one downsamples the original image using a $3\times3$ convolution with a stride of 2, while each following level is used to generate more refined  features. The final result is produced using several upsampling layers with additional skip connections introduced for feature fusion. Feature upsampling is performed with a $3\times3$ convolution followed by the pixel shuffle layer with a scale factor of 2. The model was trained to maximize the PSNR loss function on $1024\times1024$ pixel patches with a batch size of 32. Network parameters were optimized for 600 epochs using the Adam~\cite{kingma2014adam} algorithm with a learning rate of 1e--3 decreased by half every 120 epochs.

\subsection{ENERZAi}

\begin{figure}[h!]
\centering
\resizebox{1.0\linewidth}{!}
{
\includegraphics[width=0.9\linewidth]{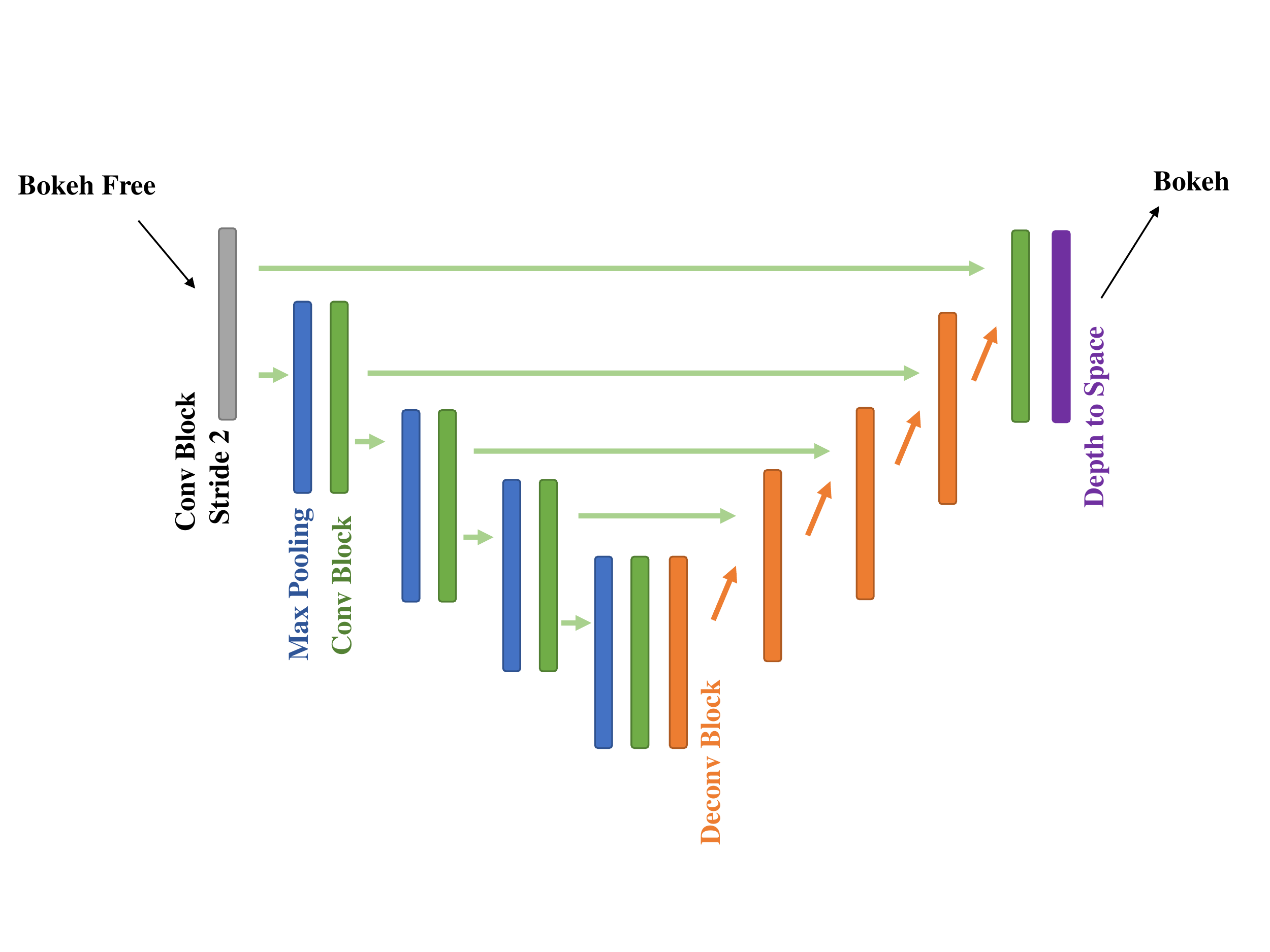}
}
\caption{\small{U-Net based model proposed by team ENERZAi.}}
\label{fig:ENERZAi}
\end{figure}

The model architecture proposed by team ENERZAi is also based on the U-Net design (Fig.~\ref{fig:ENERZAi}) that can effectively extract high-level image features. Leaky ReLU with a slope of 0.2 is used as an activation function in all convolution and deconvolution blocks, the depth-to-space op is applied at the end of the model to produce the final image output. The model was trained with a combination of the L1, multiscale-ssim (MS-SSIM) and ResNet-50 perceptual loss functions. In addition, the authors used a differentiable approximate histogram feature loss inspired by the Histogan~\cite{afifi2021histogan} to consider the color distribution of the constructed RGB image. For each R, G, and B channel, the histogram difference between the constructed and target images was calculated and added to the overall loss function. Network parameters were optimized with a batch size of 1 using the Adam algorithm with a learning rate of 1e--3 halved every 4 epochs.

\subsection{ZJUT-Vision}

\begin{figure}[t!]
\centering
\resizebox{0.9\linewidth}{!}
{
\includegraphics[width=1.0\linewidth]{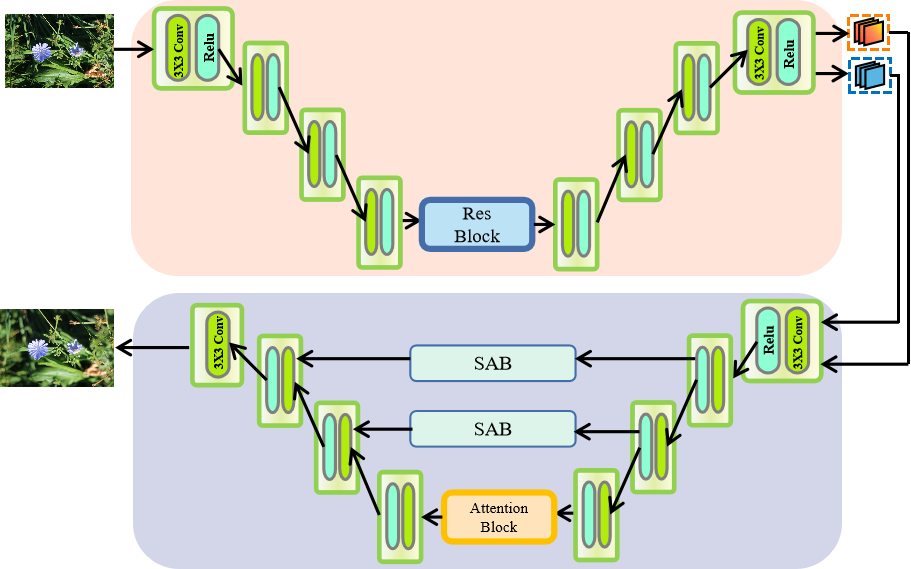}
}
\caption{\small{Attention-based Double V-Net proposed by team ZJUT.}}
\label{fig:ZJUT}
\end{figure}

The solution proposed by team ZJUT-Vision is based on the BGGAN~\cite{qian2020bggan} architecture that uses two consecutive U-Net models (Fig.~\ref{fig:ZJUT}) and two adversarial discriminators. Each U-net generator model contains nine residual blocks, and transposed convolution block is used instead of direct bilinear upsampling to improve the performance of our network. The second U-net model has also two spatial attention and one channel attention block for better visual results. Two discriminators with a different numbers of downsampling conv blocks (5 and 7) are used to extract both local and global information from the image. The authors used the Wasserstein GAN training strategy for optimizing the model. The total loss consists two components: the generator loss G combining the adversarial, perceptual, L1 and SSIM losses, and the discriminator loss D.

\subsection{Sensebrain}

The solution~\cite{lee2022bokeh} proposed by team Sensebrain consists of three main parts:

\begin{enumerate}
    \item A Dense Prediction Transformer (DPT)~\cite{ranftl2021vision} based depth estimation module that estimates the relative depth of the scene;
    \item A Nonlinear Activation Free Network (NAFNet) based generator module~\cite{chen2022simple} that takes the input image together with the predicted depth map, and outputs the depth-aware bokeh image;
    \item A dual-receptive field patch-GAN discriminator based on~\cite{qian2020bggan}.
\end{enumerate}

To generate the bokeh output, the input image is first processed by the depth prediction module to produce a depth map $\textbf{D}$. The output $\textbf{D}$ is then concatenated with the original input image $\textbf{I}$ and then passed to the NAFNet generator $\textbf{G}$ to produce the final bokeh image $\textbf{G}(\textbf{I} \odot \textbf{D})$.

The depth map of the scenes is estimated from a single monocular image using the dense prediction transformer (DPT)~\cite{ranftl2021vision} that was pre-trained on a large-scale mixed RGB-to-depth ``MIX 5'' dataset~\cite{ranftl2020towards}. The authors found empirically that this network can generate realistic depth maps with smooth boundaries and fine-grain details at object boundaries, which provides extra blurring clues to assist the backbone generator in creating a better bokeh blur. The predicted depth map is also used as a greyscale saliency mask needed for separating the background and foreground when calculating the Bokeh Loss described below.

The authors chose an 8-block NAFNet~\cite{chen2022simple} model as a baseline generator network as it demonstrated a good balance between the model complexity and performance. The core idea of the NAFNet is that the traditional high-complexity non-linear operators such as Softmax, GELU, and Sigmoid, can be removed or replaced with multiplications, which is especially useful in reducing the complexity in computationally expensive attention estimation. The encoder size was set to [2,2,2,20], the decoder size~-- to [2,2,2,2], and two NAF blocks were concatenated in the middle.

\begin{figure}[t!]
\centering
\resizebox{1.0\linewidth}{!}
{
\includegraphics[width=1.0\linewidth]{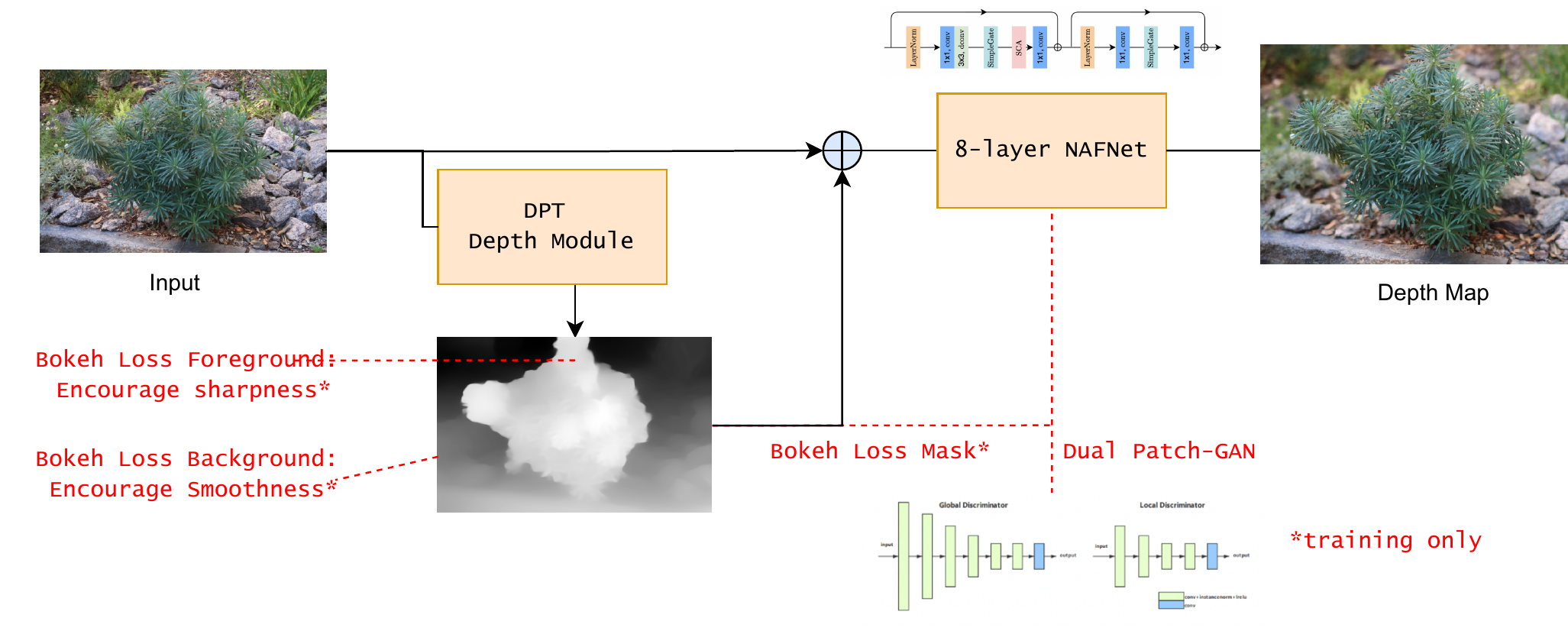}
}
\caption{\small{The overall model design proposed by team Sensebrain.}}
\label{fig:Sensebrain}
\end{figure}

The authors used adversarial learning to improve the perceptual quality of the reconstructed images. Similar to~\cite{qian2020bggan}, a multi-receptive-field Patch-GAN discriminator~\cite{isola2017image} was used as a part of the final loss function. The model takes two patch-GAN discriminators, one of depth $3$ and one of depth $5$, and averages the adversarial losses from both discriminators. Besides the conventional L1, SSIM, and discriminator loss functions, the authors also proposed a bokeh-specific loss. The core idea of this loss is that, for a natural bokeh image, the in-focus area should be sharp and the out-of-focus area should be blurred. Using the depth map we generate by the DPT module as a greyscale saliency mask $\textbf{M}$, the following loss functions are introduced:

\begin{itemize}
\item A \textit{Foreground Edge Loss} that encourages sharper edges separating the foreground and the background;
\item An \textit{Edge Difference Loss} that encourages similar edge intensities for the input and output image;
\item A \textit{Background Blur Loss} that encourages a smoother background with less noise.
\end{itemize}

The \textit{Foreground Edge Loss} $L_{foreedge}$ is used to maximize the intensity of the foreground edges. For this, the input image $\hat{\textbf{I}}$ is first multiplied with the greyscale saliency mask $\textbf{M}$, and then the edge map (gradients) is computed using the Sobel filter in the $x,y,xy,$ and $yx$ directions. L1-norms of the obtained gradients are then summed up, and the loss is defined as:
\begin{equation*}
L_{foreedge}(\hat{\textbf{I}}, \textbf{M}) = - \frac{\sum_{z \in \{x, y, xy, yx\}}\left\|S_{z}(\hat{\textbf{I}}\cdot \textbf{M})\right\|_1}{h_{\hat{\textbf{{I}}}}\cdot w_{\hat{\textbf{I}}}},
\end{equation*}
where $S_z$ is the Sobel convolution operator in the $z$ direction.

The \textit{Edge Difference Loss} is used to minimize the difference in edge strength between the input and the output images:
\begin{equation*}
L_{edgediff}(\textbf{I}, \hat{\textbf{I}}, \textbf{M}) = ||L_{foreedge}(\hat{\textbf{I}}\cdot\textbf{M})|-|L_{foreedge}(\textbf{I} \cdot \textbf{M})||.
\end{equation*}

Finally, the \textit{Background Blur Loss} $L_{backblur}$ is used to encourage a smoother blur for the background. The input image $\hat{\textbf{I}}$ is first multiplied with the inverse of the greyscale mask $\textbf{M}$, and the loss is defined as a total variation of the scene:
\begin{equation*}
L_{backblur}(\hat{\textbf{I}}, \textbf{M}) = \frac{1}{h_{\hat{\textbf{I}}}\cdot w_{\hat{\textbf{I}}}}TV(\hat{\textbf{I}} \cdot (1-\textbf{M})).
\end{equation*}

The model is trained in two stages. First, it is pre-trained without adversarial losses using the following loss function:
\begin{equation*}
L_{pretrain} = \alpha L_1 + \zeta L_{SSIM} + \kappa L_{edgediff} + \mu L_{backblur} + \nu L_{foreedge},
\end{equation*}
and during the second stage it is trained with a combination of the L1, VGG-based, SSIM and adversarial losses:
\begin{equation*}
L_{refinement}=\alpha L_1 + \beta L_{VGG} + \zeta L_{SSIM} + \lambda L_{adv},
\end{equation*}
where $\alpha = 0.5, \beta = 0.1, \zeta = 0.05, \lambda = 1, \kappa = 0.005, \mu = 0.1, \nu = 0.005$. The model was trained using the Adam optimizer with a learning rate of 1e--4 and a batch sizeof 2. The $\lambda$ value used for the WGAN-GP gradient penalty was set to 1. The dataset was additionally pruned manually to remove images that had problems with lighting and/or alignment, the final size of the training dataset was around 4225 images.

\subsection{VIC}

\begin{figure}[h!]
\centering
\resizebox{0.9\linewidth}{!}
{
\includegraphics[width=1.0\linewidth]{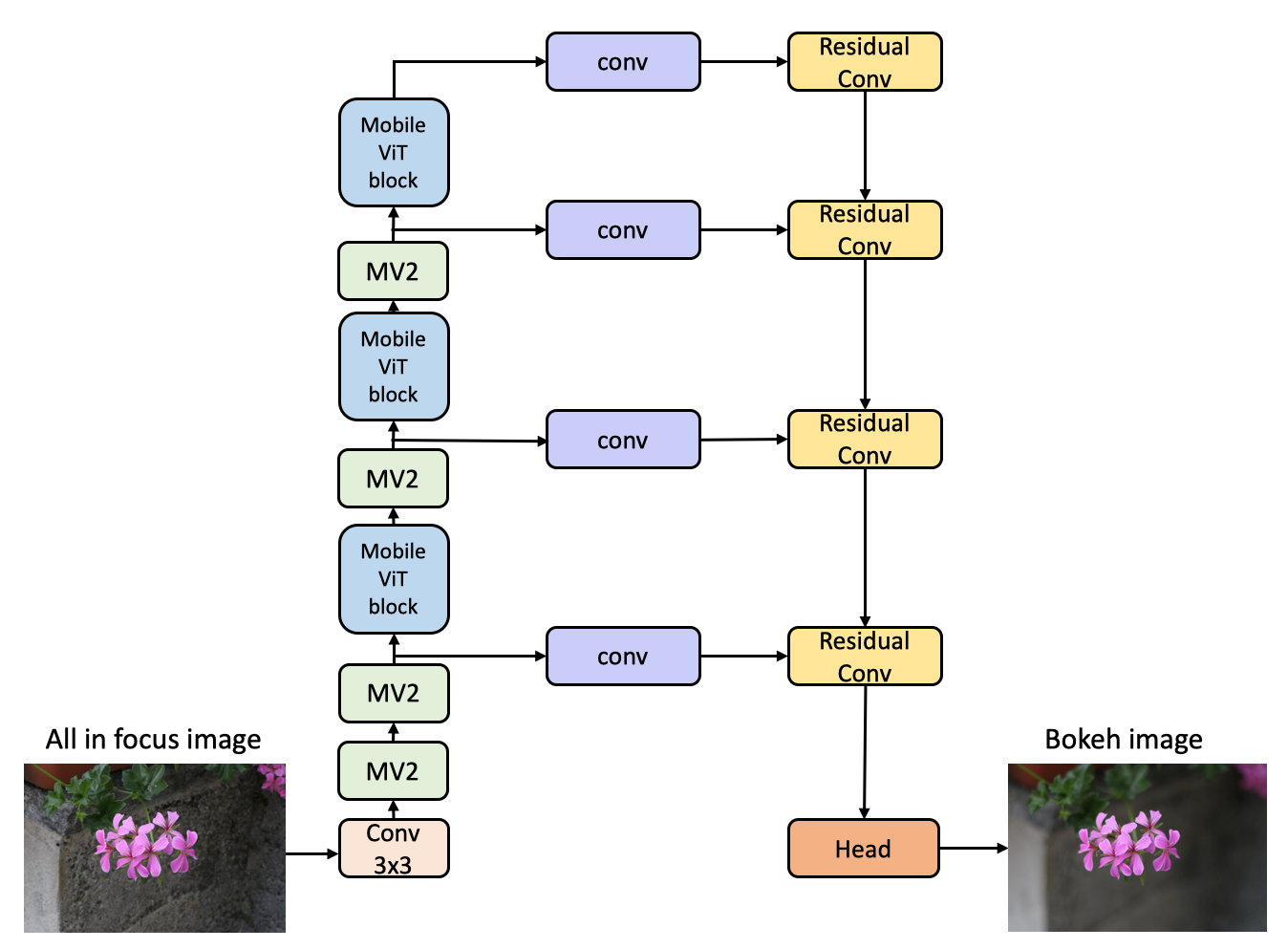}
}
\caption{\small{The diagram of the model proposed by team VIC. MV2 represents multiple MovileNetv2 blocks~\cite{sandler2018mobilenetv2}.}}
\label{fig:VIC}
\end{figure}

The overall model architecture proposed by team VIC is similar to the Dense Prediction Transformer (DPT)~\cite{ranftl2021vision} and is illustrated in Fig.~\ref{fig:VIC}. For downsampling, the encoder module uses $3\times3$ strided convolutions and MobileNetv2~\cite{sandler2018mobilenetv2} blocks.  To reduce the size of the model and its runtime, the authors used MobileViTv2~\cite{mehta2022separable} blocks in the encoder. Among various versions of the MobileViTv2, the largest one was used to improve the performance. The decoder module reassembles the features corresponding to different resolutions to produce the final bokeh image. Since there are 3 MobileViT blocks in the encoder, the decoder uses 4 input feature maps: an input of the first MobileViT block and the outputs of each MobileViT blocks.

The model was first pre-trained to learn the identity mapping during the first 10 epochs with the L1 loss function. Next, it was trained for another 10 epochs with the L1 loss to perform the considered bokeh effect rendering task. The authors also used the mask made from the DPT~\cite{ranftl2021vision} disparity map at the end of the training: since this mask can roughly distinguish the foreground and the background, one can apply the L1 loss between the foreground of the predicted image and the foreground of the original all-in-focus image. By doing this, the model can learn how to produce distinct foreground objects. To learn the blur, L1 loss between the background of a predicted image and the background of the ground-truth bokeh image was applied. Feature reconstruction loss~\cite{johnson2016perceptual} was also added without using the mask. The model was trained with a batch size of 2 on images resized to $512\times768$ pixels and augmented with random flips and rotations. Network parameters were optimized using the AdamW algorithm with an initial learning rate of 5e--4 reduced to 1e--6 with the cosine scheduler.

\subsection{MiAIgo}

\begin{figure}[h!]
\centering
\resizebox{1.0\linewidth}{!}
{
\includegraphics[width=0.9\linewidth]{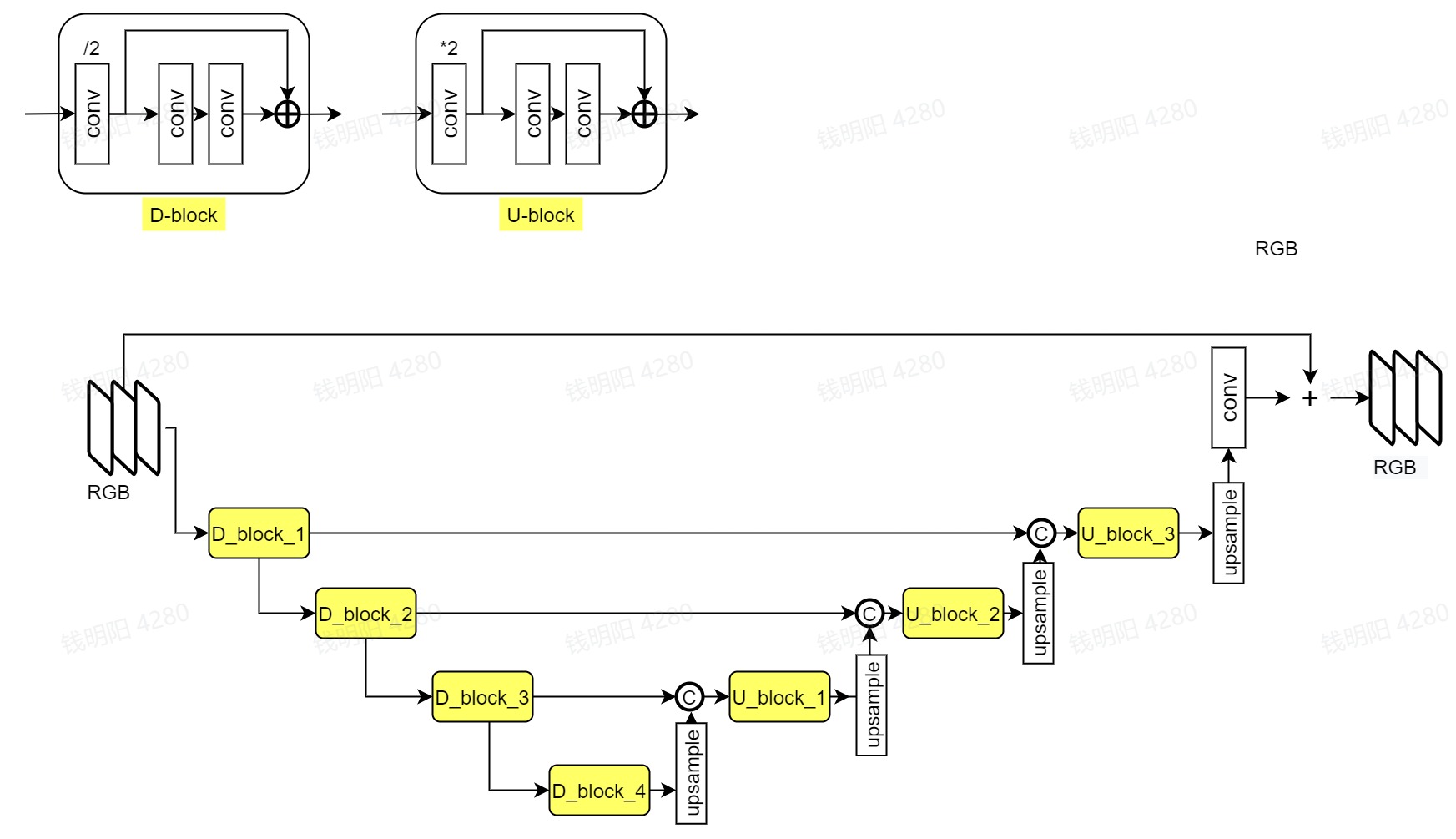}
}
\caption{\small{Network architecture developed by team MiAIgo.}}
\label{fig:MiAIgo}
\end{figure}

Team MiAIgo used an encoder-decoder model design for fast feature extraction (Fig.~\ref{fig:MiAIgo}) that produces a high-resolution disparity map used to generate realistic bokeh rendering effect. First, the input image is resized to 1024${\times}$1024 px and then processed by seven model blocks that are extracting multi-scale contextual information and generating a high-resolution disparity map. Each block consists of three convolution layers with one additional skip connection. At the end of the decoder module, the obtained disparity map is superimposed with the original image to get the final bokeh output. The model was trained with a combination of the L1, L2 and SSIM loss functions. Network parameters were optimized using the Adam algorithm with a learning rate of 5e--5 and a batch size of 8.

\section{Additional Literature}

An overview of the past challenges on mobile-related tasks together with the proposed solutions can be found in the following papers:

\begin{itemize}
\item Learned End-to-End ISP:\, \cite{ignatov2019aim,ignatov2020aim,ignatov2022microisp,ignatov2022pynetv2}
\item Perceptual Image Enhancement:\, \cite{ignatov2018pirm,ignatov2019ntire}
\item Bokeh Effect Rendering:\, \cite{ignatov2019aimBokeh,ignatov2020aimBokeh}
\item Image Super-Resolution:\, \cite{ignatov2018pirm,lugmayr2020ntire,cai2019ntire,timofte2018ntire}
\end{itemize}

\section*{Acknowledgements}

We thank the sponsors of the Mobile AI and AIM 2022 workshops and challenges: AI Witchlabs, MediaTek, Huawei, Reality Labs, OPPO, Synaptics, Raspberry Pi, ETH Z\"urich (Computer Vision Lab) and University of W\"urzburg (Computer Vision Lab).

\appendix
\section{Teams and Affiliations}
\label{sec:apd:team}

\bigskip

\subsection*{Mobile AI 2022 Team}
\noindent\textit{\textbf{Title: }}\\ Mobile AI 2022 Learned Smartphone ISP Challenge\\
\noindent\textit{\textbf{Members:}}\\ Andrey Ignatov$^{1,2}$ \textit{(andrey@vision.ee.ethz.ch)}, Radu Timofte$^{1,2,3}$\\
\noindent\textit{\textbf{Affiliations: }}\\
$^1$ Computer Vision Lab, ETH Zurich, Switzerland\\
$^2$ AI Witchlabs, Switzerland\\
$^3$ University of Wuerzburg, Germany\\

\subsection*{Antins\_cv}
\noindent\textit{\textbf{Title:}}\\ A Tiny UNet for Image Bokeh Rendering \\
\noindent\textit{\textbf{Members: }}\\ \textit{Jin Zhang (zj346862@antgroup.com)}, Feng Zhang, Gaocheng Yu, Zhe Ma, Hongbin Wang\\
\noindent\textit{\textbf{Affiliations: }}\\
Ant Group, China\\

\subsection*{ENERZAi}
\noindent\textit{\textbf{Title:}}\\ Bokeh Unet and Histogram Feature Loss \\
\noindent\textit{\textbf{Members: }}\\ \textit{Minsu Kwon (minsu.kwon@enerzai.com)}\\
\noindent\textit{\textbf{Affiliations: }}\\
ENERZAi, South Korea\\

\subsection*{ZJUT-Vision}
\noindent\textit{\textbf{Title:}}\\ Attention-based Double V-Net\\
\noindent\textit{\textbf{Members: }}\\ \textit{Haotian Qian (1092944263@qq.com)}, Wentao Tong, Pan Mu, Ziping Wang, Guangjing Yan\\
\noindent\textit{\textbf{Affiliations: }}\\
Zhejiang University of Technology, China\\

\subsection*{Sensebrain}
\noindent\textit{\textbf{Title:}}\\ Bokeh-Loss GAN: Multi-stage Adversarial Training for Realistic Edge-Aware Bokeh~\cite{lee2022bokeh} \\
\noindent\textit{\textbf{Members: }}\\ \textit{Brian Lee$^1$ (brianlee@sensebrain.site)}, Lei Fei$^2$, Huaijin Chen$^1$ \\
\noindent\textit{\textbf{Affiliations: }}\\
$^1$ Sensebrain Technology, United States\\
$^2$ Tetras.AI, China\\

\subsection*{VIC}
\noindent\textit{\textbf{Title:}}\\ Mobile-DPT for Bokeh Rendering\\
\noindent\textit{\textbf{Members: }}\\ \textit{Hyebin Cho (jhb0316@kaist.ac.kr)}, Byeongjun Kwon, Munchurl Kim\\
\noindent\textit{\textbf{Affiliations: }}\\
Korea Advanced Institute of Science and Technology (KAIST), South Korea\\

\subsection*{MiAIgo}
\noindent\textit{\textbf{Title:}}\\ Realistic Bokeh Rendering Model on Mobile Devices\\
\noindent\textit{\textbf{Members: }}\\ \textit{Mingyang Qian (qianmingyang@xiaomi.com)}, Huixin Ma, Yanan Li, Xiaotao Wang, Lei Lei\\
\noindent\textit{\textbf{Affiliations: }}\\
Xiaomi Inc., China\\

{\small
\bibliographystyle{splncs04}
\bibliography{egbib}
}

\end{document}